\documentclass[12pt,twocolumn]{aastex62}

\begin{document}

\title{Trans-Neptunian Objects Transiently Stuck in Neptune's Mean Motion Resonances: Numerical simulations of the current population}

\author[0000-0002-5228-7176]{Tze Yeung Mathew Yu}
\affiliation{Department of Physics and Astronomy, University of California, Los Angeles}

\author[0000-0001-5061-0462]{Ruth Murray-Clay}
\affiliation{Department of Astronomy and Astrophysics, University of California, Santa Cruz, 1156 High St, Santa Cruz, CA 95064, United States}

\author[0000-0001-8736-236X]{Kathryn Volk}
\affiliation{Lunar and Planetary Laboratory, University of Arizona, 1629 E University Blvd, Tucson, AZ 85721, United States}

\begin{abstract}
A substantial fraction of our solar system's trans-Neptunian objects (TNOs) are in mean motion resonance with Neptune. Many of these objects were likely caught into resonances by planetary migration---either smooth or stochastic---approximately 4 Gyr ago. Some, however, gravitationally scattered off of Neptune and became transiently stuck in more recent events. Here, we use numerical simulations to predict the number of transiently-stuck objects, captured from the current actively scattering population, that occupy 111 resonances at semimajor axes $a=$30--100~au. Our source population is an observationally constrained model of the currently-scattering TNOs. We predict that, integrated across all resonances at these distances, the current transient sticking population comprises 40\% of total transiently-stuck+scattering TNOs, suggesting that these objects should be treated as a single population. We compute the relative distribution of transiently-stuck objects across all $p$:$q$ resonances with $1/6~\le~q/p~<~1$, $p<40$, and $q<20$, providing predictions for the population of transient objects with $H_r < 8.66$ in each resonance. We find that the relative populations are approximately proportional to each resonance's libration period and confirm that the importance of transient sticking increases with semimajor axis in the studied range. We calculate the expected distribution of libration amplitudes for stuck objects and demonstrate that observational constraints indicate that both the total number and the amplitude-distribution of 5:2 resonant TNOs are inconsistent with a population dominated by transient sticking from the current scattering disk. The 5:2 resonance hence poses a challenge for leading theories of Kuiper belt sculpting.
\end{abstract}

\keywords{Kuiper belt: general}

\section{Introduction}\label{s:intro}

Trans-Neptunian objects (TNOs) in mean motion resonance with Neptune provide a unique probe of the solar system's early dynamical history.  These objects have been interpreted as evidence of Neptune's early migration due to smooth transfer of angular momentum with a remnant planetesimal disk \citep[e.g.,][]{Malhotra1993,Malhotra1995,Hahn2005}, high-eccentricity scattering following a dynamical upheaval event \citep[e.g.,][]{Levison2008}, or a combination of these processes (reviewed in \citealt{Morbidelli2008}; see also \citealt{Dawson2012}).  However, transient sticking---a resonance capture mechanism \textit{not} associated with planetary migration---contributes to the Kuiper belt's resonant population as well \citep[e.g.,][]{Levison1997,Lykawka2007,Gomes2008}. In this work, we model the population of resonant TNOs produced by transient sticking from the solar system's current ``scattering disk." 

Transient sticking occurs when members of the scattering disk---objects undergoing repeated scatterings by Neptune---are temporarily caught into mean motion resonance with the planet.  Repeated scattering causes the semimajor axes and eccentricities of scattering disk objects to random walk.  Because this evolution approximately maintains the objects' pericenter distance (corresponding to regions near the semimajor axis of Neptune, where scatterings occur), on average it increases a TNO's semimajor axis until it is distant enough to be detached by galactic tides, joining the Oort cloud \citep[e.g.,][]{Dones2004,Gomes2008}. When objects random walk to semimajor axes corresponding to mean-motion resonances, they often experience temporary libration within resonance.  Because this transient sticking rarely produces objects that are tightly bound within resonances, it is not thought to be the primary production mechanism for the most studied populations of resonant TNOs---those in the 3:2 and 2:1 resonances.  However, the recent confirmation of an unexpectedly large population of 5:2 resonant objects \citep{Gladman2012,Volk2016}, as well as the detection of objects known to be transient members of Neptune's distant resonances \citep{Bannister201692,Holman2018}, brings new urgency to the characterization of the transiently-stuck population.

Neither smooth migration nor dynamical upheaval models, in their standard forms, predict a large population in the 5:2 resonance \citep[e.g.,][]{Chiang2003,Hahn2005,Levison2008}, leaving transient sticking as an alternative standard theory available to explain their presence.  In this work, we use numerical simulations to model the population of objects produced by transient sticking in the Kuiper belt's current actively scattering population. We note that because some transient sticking events are in fact long-lived, today's resonances may contain a population of objects that were transiently stuck early in the solar system's lifetime, when the orbital distribution of the scattering-object source population may have differed from the population today. We do not model this interesting possibility here, though we refer the reader to several recent studies that include scattering and resonance capture during Neptune's early epoch of migration \citep{Kaib2016,Nesvorny2016,Pike2017}.  We note that \citet{Pike2017} are able to reproduce a large 5:2 population with a dynamical upheaval model, though not all resonance populations in their model match observational constraints.  

In this paper, our dual goals are to (1) characterize the population of transiently-stuck TNOs that should be excluded from studies of primordial TNO dynamics and (2) determine whether the 5:2 population is consistent with exclusive emplacement by transient sticking of the {\it current} scattering population.  
In Section~\ref{s:methods} we outline the numerical simulations used to investigate resonance sticking in the current population of scattering TNOs.  Section~\ref{s:results} outlines the major results from these simulations.  We discuss these results in light of observations in Section~\ref{sec:discussion} and provide a summary of our main conclusions in Section~\ref{s:summary}.

\section{Methods}\label{s:methods}

To determine the orbital distribution of currently resonant objects that may be attributed to transient sticking from today's population of scattering TNOs, we perform a series of numerical simulations.  
We outline our initial conditions in Section~\ref{ss:sims}. Section~\ref{ss:resid} describes how simulation output was searched for instances of temporary sticking in Neptune's mean motion resonances.  We demonstrate consistency across our simulations and summarize simulation statistics in Section~\ref{ss:datamp}.

\subsection{Initial Conditions: a model of the current scattering population}\label{ss:sims}

We use a model of the current scattering population (see \citealt{Gladman2008} for a detailed definition of this population) for our initial conditions. Following \citet{Alexandersen2013} and \citet{Shankman2013}, we use initial conditions from a scattering population simulated in \citet{Kaib2011}, with inclinations adjusted to reflect a dynamically hotter initial disk of particles. This is necessary because the original \citet{Kaib2011} simulations produced a scattering population with inclinations too low to match observational constraints \citep{Shankman2013}. The initial inclination distribution for the actively scattering particles from the modified  \citet{Kaib2011} model is similar to an offset Gaussian (such as that used by \citep{Gulbis2010}) with a mean of $\sim16^{\circ}$ and a width of $\sim7^{\circ}$. Our initial conditions for the simulations consist of 8500 TNO particles; only particles whose semimajor axes changed by 1.5~au or more within the last 10~Myr of the \citet{Kaib2011} model run are included, as this is how the observed actively scattering TNO population is defined.
\citet{Shankman2016} and \citet{Lawler2018} found that these initial conditions provide an adequate representation of the population of scattering TNOs observed by the Canada France Ecliptic Plane Survey \citep[CFEPS][]{Petit2011}, by the Outer Solar System Origins Survey \citep[OSSOS][]{Bannister2016,Bannister2018}, and by \citet{Alexandersen2016}.
We numerically integrate the orbits of the model scattering population as massless test particles using the rmvs3 routine in SWIFT \citep{Levison1994} with the Sun and the four giant planets included as massive bodies. Test particles remain in the simulation until they collide with a planet or until they reach a heliocentric distance interior to Jupiter or farther than 1000~au from the Sun. We only record and analyze test particle behavior during their residence in the semimajor axis range $30<a<100$~au, so the inner and outer boundaries of the simulations should not affect our results.

\begin{figure}[htbp]
   \centering
   \includegraphics[width=3in]{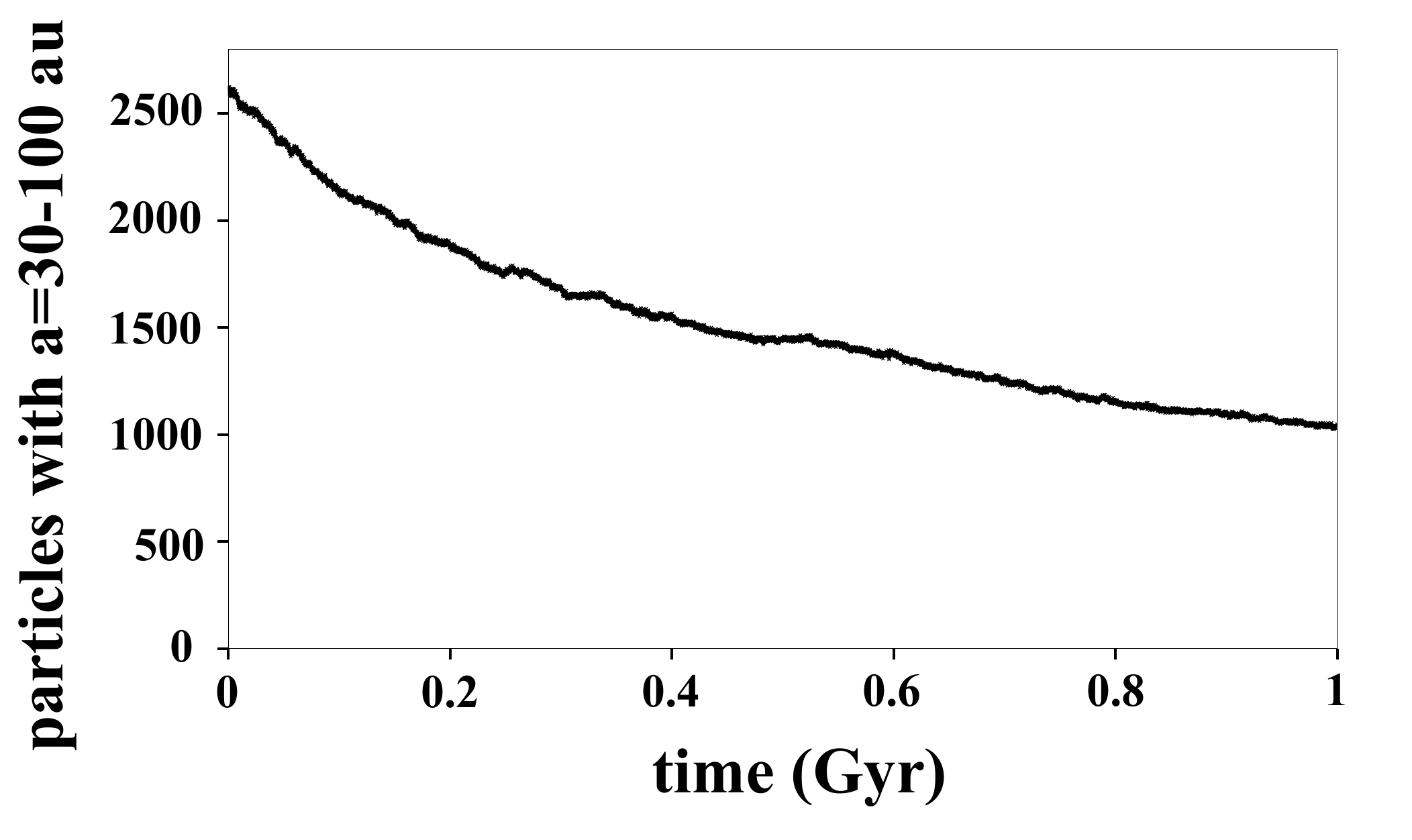}
   \caption{The number of test particles with $30<a<100$~au as a function of time in our 1~Gyr simulation. The number declines to $\approx$40\% of its initial value over 1~Gyr. }\label{f:particle_in_time}
\end{figure}

We are interested in resonance sticking at all timescales, from $10^5$ years (a few libration periods for close-in resonances) to the age of the solar system.  However, the total population of scattering TNOs has decayed over time.  Figure~\ref{f:particle_in_time} displays the number of particles with $30<a<100$~au in a simulation lasting 1~Gyr.  Over this timescale, the population declines to $\approx$40\% of its initial value.  We thus restrict ourselves to $10^9$-year timescales and treat any remnant resonant TNOs that were ``transiently" stuck more than 1~Gyr ago as part of the primordial (i.e., not modeled) population.  
We note that over the course of the 1~Gyr simulation, the peak inclination of the test particles in the $a=30-100$~au range shifts up by $\sim$$1^\circ$, with the width of the distribution also increasing by $\sim$$1^\circ$; the distribution of pericenter distances also shifts, with the peak increasing from 33.7~au to 35.2~au. We consider these changes acceptable.

Ideally, we would perform a single, high-time-resolution simulation covering all resonance sticking timescales ($10^5-10^9$ years).
 However, as we discuss in Section~\ref{ss:resid}, reliably identifying libration within a resonance requires a minimum of $\sim$$1000$ time points; saving output from a $10^9$ year simulation at a resolution of $10^3$ years is very resource intensive both in terms of required disk space and cpu time for data analysis. We thus compromised and performed three separate numerical simulations from the same initial conditions to cover three different total time spans ($1.5\times 10^7$, $10^8$, and $10^9$ years) at different output frequencies. These are used to investigate resonance sticking at three timescale ranges: $10^5-1.5\times 10^7$ years, $10^6-10^8$ years, and $10^7-10^9$ years.  Section~\ref{ss:datamp} describes how we normalize and combine the results of these simulations to analyze the entire range of resonance sticking timescales we wish to investigate.

\subsection{Identifying resonances in the simulations}\label{ss:resid}

A mean motion resonance with Neptune occurs when a TNO's orbital period is related to Neptune's orbital period by a simple integer ratio; this relationship between orbital periods ensures that conjunctions between the TNO and Neptune are repeated at specific geometric arrangements, creating a resonant perturbation on the TNO's orbit. We designate Neptune's exterior resonances as $p$:$q$ resonances, where $p$ and $q$ are integers~$\ge~1$ with $p~>~q$.  Each $p$:$q$ resonance has several possible arguments in the disturbing potential that could affect a TNO's orbital evolution \citep[see, e.g.,][]{MurrayDermott1999}. Scattering TNOs typically have large eccentricities, so the strongest resonant argument for a given $p$:$q$ mean motion resonance is the one that involves the TNO's eccentricity ($e_{tno}$) and longitude of perihelion ($\varpi_{tno}$). These arguments have a resonant angle, $\phi$, that takes the form:
\begin{equation}\label{eq:phi}
\phi = p \lambda_{tno} - q \lambda_{N} - (p-q)\varpi_{tno}
\end{equation}
where $\lambda_{tno}$ and $\lambda_{N}$ are the mean longitudes of the TNO and Neptune. When an object is in resonance, $\phi$ will librate around some stable value (usually, but not always, $180^{\circ}$) with an amplitude $A_{\phi}$ (which we define as half the peak-to-peak amplitude); objects far from resonance will have quickly varying values of $\phi$ that explore the full range from $0-360^{\circ}$. We checked for libration of $\phi$ for all resonances with values of $p<40$ and $q<20$ in the period ratio range $1/6\le~q/p~<1$, which corresponds to the semimajor axis range $30<a\le100$~au given Neptune's semimajor axis $a_N \approx 30$~au.

The goal of analyzing our numerical simulation data is to determine if, when, and how long a test particle sticks to (i.e., experiences libration within) any of these mean motion resonances. 
Because transiently-stuck TNOs do not occupy resonances for the full length of the simulation, we search for sticks using a series of sliding windows in time.
To avoid spurious resonance identification, we require identified sticks to contain at least 1000 data points, meaning that the minimum identifiable stick length is a factor of $10^3$ longer than the simulation output's time resolution.
For the $1.5\times10^7$~year simulation, we output orbital information every $10^2$ years, so the minimum window length is $10^5$ years. Our $10^8$ and 10$^9$ year simulations had output every $10^3$ and $10^4$ years, corresponding to minimum windows of $10^6$ and $10^7$ years, respectively.

We search for sticks using these running time windows as follows. First, we check to see if a test particle maintains a constant period ratio with Neptune during the window being considered.
The instantaneous period ratio between a TNO and Neptune is $f_P \equiv (a_{tno}/a_N)^{3/2}$, where $a_{tno}$ is the particle's osculating semimajor axis.  If a test particle experiences significant changes in orbital period (or, equivalently, semimajor axis) within a time window, then the window is discarded as non-resonant because the particle is actively gravitationally scattering; our conservative threshold for a window to be considered non-resonant due to scattering is defined as period ratio change $\Delta f_P \ge 0.2$.
If a test particle's evolution within a window does not display obvious scattering behavior, we use the period ratio to determine which possible resonances to check for.
We consider a particle to be close to a predicted resonance if $\bar f_P \equiv (\bar a_{tno}/a_{N})^{3/2} = q/p \pm 0.1$, where $\bar a_{tno}$ is the mean semimajor axis over the time window;   this typically means $\bar a_{tno}$ is within $\sim2.5$~au of the nominal resonant semimajor axis. For resonant ratios meeting this requirement, we calculate the corresponding resonance angle (Equation~\ref{eq:phi}) over the time window and check for libration. The initial calculation of $\phi$ is done such that $\phi$ falls in the range $[0,360^{\circ})$.

\begin{figure}[htb]
   \centering
   \begin{tabular}{c}
	   \includegraphics[width=3in]{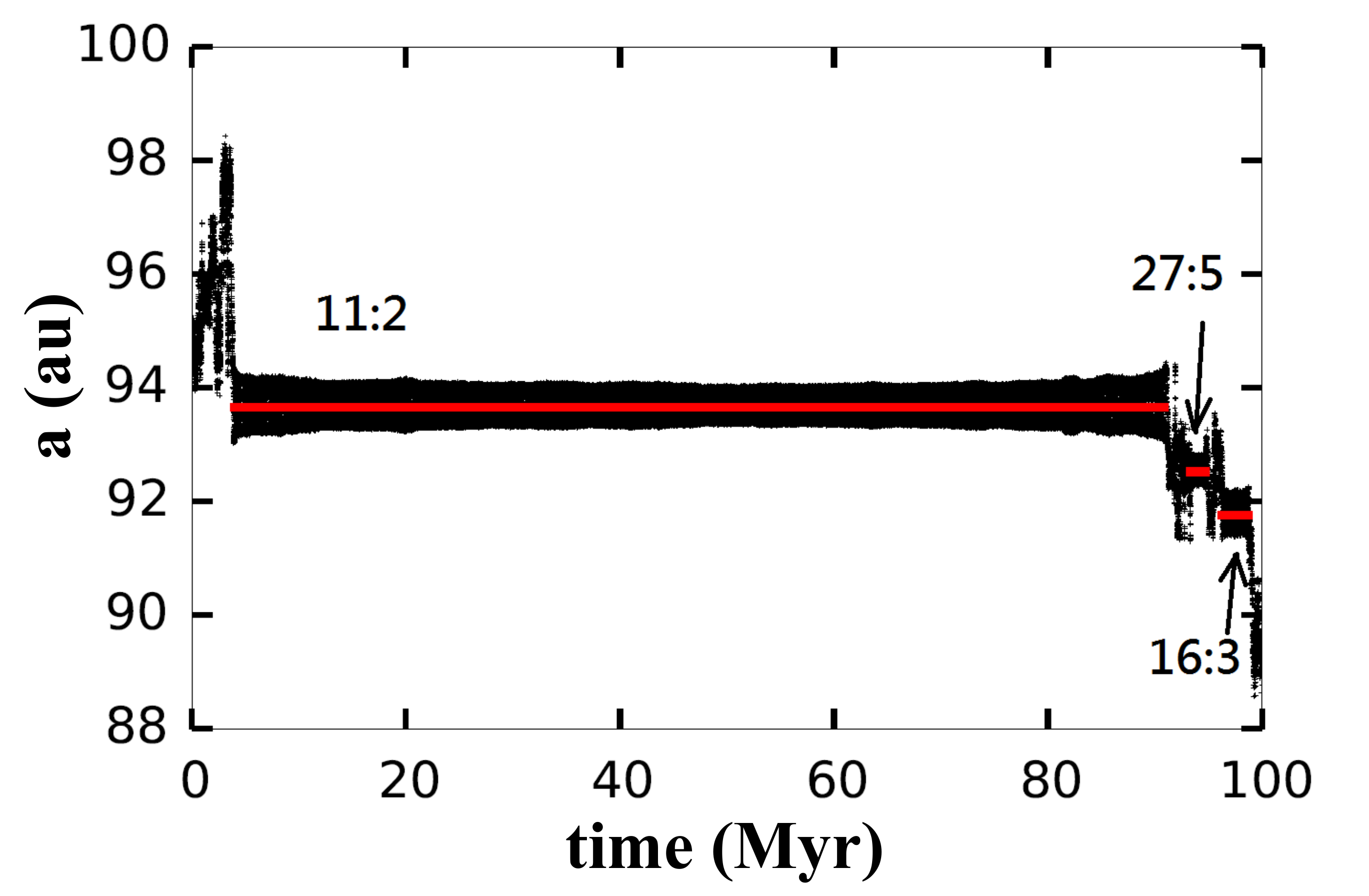}\\
	   \includegraphics[width=3in]{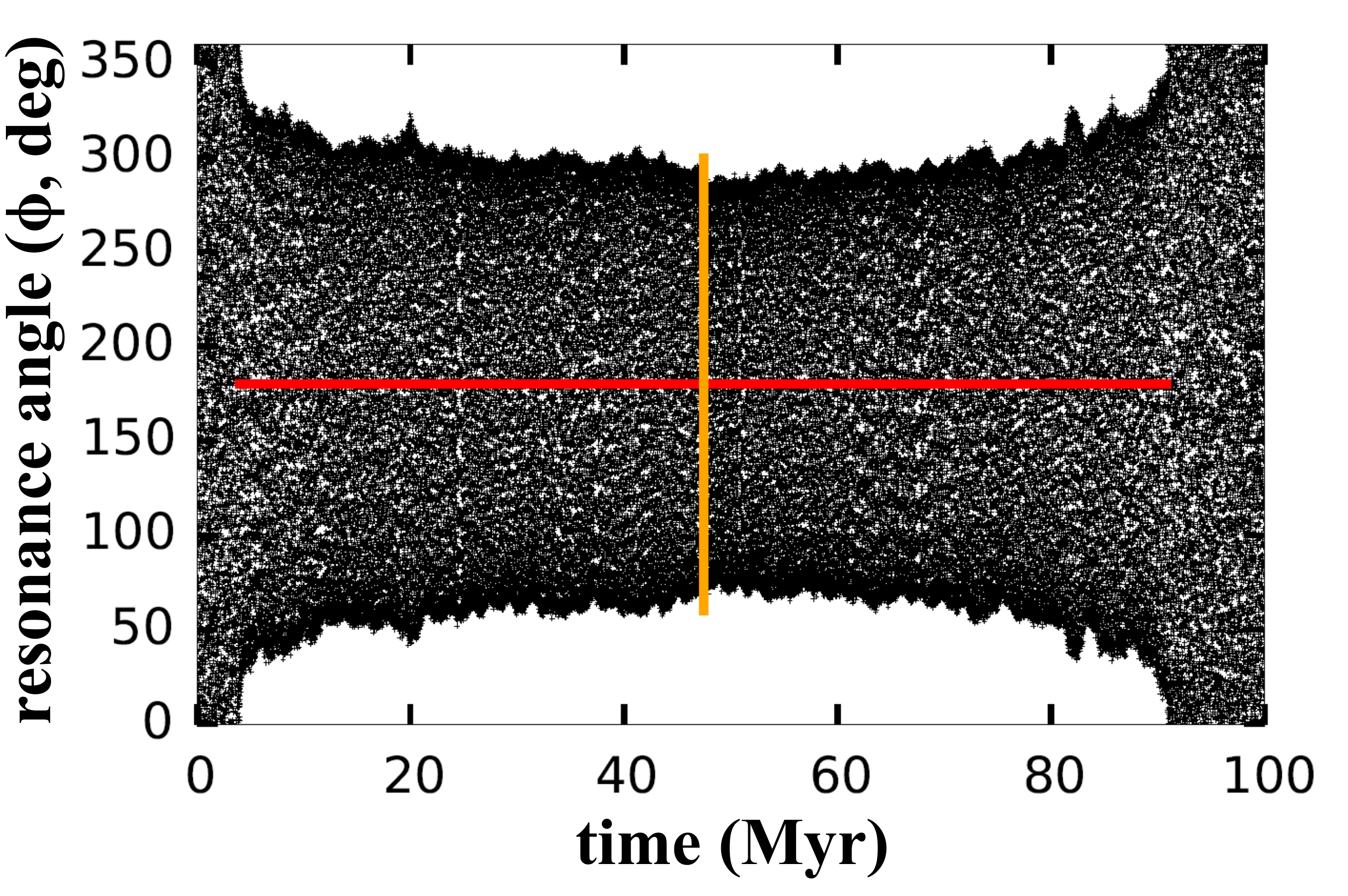}\\
   \end{tabular}
   \caption{Top: evolution of a test particle in the 100~Myr simulation. The semimajor axis evolution shows three periods of resonance sticking (red lines). Bottom: resonance angle evolution for the 11:2 stick identified in the top panel. The horizontal red line marks the duration of the stick as identified by our analysis code and the vertical orange line indicates the identified libration amplitude.}\label{f:example}
\end{figure}

To check for libration in a time window, we calculate a provisional libration amplitude defined as half the difference between the mean of the 10 largest $\phi$ values the mean of the 10 smallest $\phi$ values.
If this provisional libration amplitude is larger than $175^{\circ}$, the window is discarded as non-resonant; a $175^{\circ}$ threshold provided the best match between manual and automated resonance identification for a subset of particles which were examined by eye.
For windows with amplitudes consistent with libration, we record the time and which resonance showed libration.
Once a window has been either discarded or recorded as resonant, the window slides forward along a test particle's time series by 10 data points, and the analysis is repeated.
This process continues until the last data point for a test particle is processed. After all windows for a test particle have been processed, overlapping windows showing libration within the same resonance are linked together to identify continuous sticks.
This analysis results in a list of all the resonances a test particle stuck to during its evolution along with the the starting and ending times of the libration within each resonance.
Figure~\ref{f:example} shows a test particle from the $10^8$-year simulation with the three identified resonance sticking events labeled in the top panel; the bottom panel shows the resonance angle for the longest-lived sticking event.

For each identified resonance sticking event, we recalculate the test particle's libration amplitude.  The libration amplitude calculated during the resonance identification process is not always accurate because the individual windows might represent a small portion of the total sticking time. For sticks that cover at least 2000 data points (twice the length of an individual window in the identification process), we take the entire time period for the stick and divide it into 20 sections. Within each section, we average the minimum 2\% and the maximum 2\% of the $\phi$ values and take half the difference between these averages to be the libration amplitude for that section. We then take the median amplitude across all 20 sections to be the overall libration amplitude for that individual resonance sticking event. For resonances that contain fewer than 2000 data points, the process is the same except that the resonance is only divided into 6 sections; a finer division would result in arbitrarily smaller libration amplitudes because there would not be a sufficient number of data points within each section to accurately determine the minimum and maximum values of $\phi$.  Figure~\ref{f:example} shows an example libration amplitude calculated using this method.

The above procedure readily identifies sticking events and libration amplitudes for well-behaved libration that is centered about $\phi=180^{\circ}$, which is the expected center for stable libration in most of Neptune's resonances.
Neptune's $p$:1 resonances allow libration about so-called asymmetric libration centers \citep[see, e.g.,][]{Beauge1994}.
The exact centers for asymmetric libration depend on eccentricity \citep[see, e.g.][]{Nesvorny2001}, but typically $\phi$ librates about $\sim70^{\circ}$ or $\sim290^{\circ}$.
Most asymmetric librators are readily identified using the same procedure as that outlined above for the symmetric librators.
However, in some cases of asymmetric libration, $\phi$ passes through $0^{\circ}$ over the course of its libration; the procedure outlined above would not identify the portion of the resonance stick that included passing through $\phi=0$.
Because $\phi=0$ is generally an unstable point in the co-planar problem for the scattering population, such instances of libration slipping through $\phi=0$ are typically very short in duration and limited to higher inclination objects.
We examined the libration behavior for many known objects and determined that these slips through $\phi=0$ are rare and typically so short that our averaging procedure outlined above would be insensitive to them; in other cases, our procedure would merely result in the single stick being split into two separate sticks (divided in time at the point where the crossing occurs).
We note that at very high eccentricities, stable libration in resonance is theoretically possible around $\phi=0$ \citep[see,][]{Wang:2017,Malhotra:2018}; however for Neptune's resonances, this stable zone occurs at very high, deeply planet-crossing eccentricities, which should result in only very short sticks. \cite{Malhotra:2018}, for example, note an observed TNO that is experiencing a temporary stick around $\phi=0$ in Neptune's 5:2 resonance. The stick lasts only $\sim5\times10^4$ years, which is below our minimum stick length threshold. Thus we expect that such sticking events only very minimally contribute to the time-averaged resonant population.

Calculation of libration amplitudes for $p$:$1$ resonances is also affected by the presence of the asymmetric libration centers. Asymmetric libration occurs during a significant fraction of sticks in $p$:$1$ resonances.  However, because transient sticks to $p$:$1$ resonances often involve transitions between the two asymmetric islands and between symmetric and asymmetric libration, our libration amplitude procedure is not sufficiently accurate for $p$:$1$ objects.  A more accurate calculation of these amplitudes could be accomplished by splitting $p$:$1$ sticks into a series of mini-sticks to each of the libration islands, each with a separately calculated libration amplitude.  For this work, however, we simply note that the measured libration amplitude distribution for $p$:$1$ objects should be interpreted with caution.  Our procedure often identifies sticks that quickly switch between asymmetric islands as symmetric librators, and a small minority of quickly-transitioning sticks are assigned libration amplitudes intermediate between the asymmetric and symmetric values (c.f. Figure \ref{f:amp_time_count_all}).

\subsection{Simulation matching and stick statistics}\label{ss:datamp}

\begin{deluxetable*}{lll|lr|lrr}[htb!]
\tabletypesize{\footnotesize}
\tablecolumns{8}
\tablewidth{0pt}
\tablecaption{Number of simulated particles experiencing sticking at $30<a<100$~au \label{tab-sticks}}
\tablehead{
\multicolumn{3}{c}{simulation} & \multicolumn{2}{c}{unique particles}  		&  \multicolumn{3}{c}{number of sticks} \\
 	  	&     & 	minimum stick & in $a$ & with $\ge1$ 	& stick	& 	&  \\
id	 & length (yr)	& resolution (yr) & range  & stick  	&   timescale (yr)                        			& absolute	& \multicolumn{1}{r}{normalized}}
\startdata
1 	&  1.5$\times$10$^7$	& 10$^5$	& 3393	& 2465	& 10$^5$--10$^6$ 			& 22493	& 1499533		\\ 
   	&							&			& 	\phantom{3393}	& 	\phantom{3393}	& 10$^6$--10$^7$			& 1544	& 51467		\\
 	&							&			& 	\phantom{3393}	& 	\phantom{3393}	& (1--1.5)$\times10^7$	& 52	& 1156		\\		\hline		 
2 	& 10$^8$			& 10$^6$	& 4087	& 2629	& 10$^6$--10$^7$			& 10453	& 56719		\\
  	&							&			& 	\phantom{3393}	& 	\phantom{3393}	& 10$^7$--10$^8$			& 895	& 4209		\\	\hline
3 	& 10$^9$ 			& 10$^7$	& 5126	& 2310	& 10$^7$--10$^8$			& 6646	& 4886		\\
   	&						&			& 	\phantom{3393}	& 	\phantom{3393}	& 10$^8$--10$^9$			& 437	& 733		\\	    
\enddata 
\tablecomments{All simulations began with 8500 scattering test particles (see Section~\ref{ss:sims} for a description of their initial conditions). The number of unique particles in range refers to the number of those 8500 test particles that enter the range $30<a<100$~au at any time during the simulation. The number of sticks identified in each of our three simulations is given as a function of the stick timescale. The absolute number of sticks is given as well as the number normalized to account for the overall simulation length and test particles loss (described in Section \ref{ss:datamp} and shown in Figure \ref{f:simulation_matching}).
}
\end{deluxetable*}

To investigate the full range of stick timescales from $10^5-10^9$ years, we must combine our results by applying a weighting factor for each stick from each simulation.  First, we must adjust for the total length of each simulation (shorter simulations produce fewer sticks simply because they are shorter).  Second, as the resonance sticking population is constantly perturbed by the gas giants, objects tend to move out of our range of interest, and our simulated population declines over time as shown in Figure~\ref{f:particle_in_time}. The scattered population declines by $\approx$15\% in the 100~Myr and $\approx$60\% in the 1~Gyr simulations, so we must correct for the changing number of objects available for sticking.  Finally, some sticking timescales are covered in more than one simulation, and we do not want to over-count. The weighting factor thus has three components:
\begin{enumerate}
\item A factor of $1{\rm Gyr}/t_{sim}$, where $t_{sim}$ is the total length of the simulation from which the stick is drawn.  This corrects for the total simulation length.
\item A factor of $(T_{range,15Myr}/T_{range,sim})(t_{sim}/{\rm 15Myr})$, where $T_{range,sim}$ is the total amount of time spent by particles in the range 30-100au in the stick's simulation, and $T_{range,15Myr}$ is the same number for the 15~Myr simulation.  This corrects for the decline in the number of objects available for sticking.  We use the 15~Myr simulation for reference because it experiences the smallest decline, and we have scaled our total population to the total population observed today.
\item When simulations are combined, a factor of one divided by the number of simulations in which the stick's timescale is included.  This prevents double-counting.  This factor is omitted when simulations are analyzed separately.
\end{enumerate}
We note that factor number 2 is an improvement over simply multiplying by the percentage decline in the number of objects in each simulation since it takes into account the smoothly varying loss of particles over time.

Table \ref{tab-sticks} summarizes the number of sticking events found in each simulation for various stick length ranges; we list both the recorded number of sticks as well as the normalized number of sticks calculated using the method described above. The top panel in Figure~\ref{f:simulation_matching} displays the normalized number of resonance sticks as a function of stick duration.  The bottom panel in Figure~\ref{f:simulation_matching} recomputes this distribution with each stick weighted by its duration, yielding the total combined time spent by objects in resonance in each stick timescale bin.  This total simulated sticking time is proportional to the likelihood of finding an object in resonance at a given snapshot in time (e.g., today), given the, in this case reasonable, assumption that sticks may be approximated as statistically random and uncorrelated.

Statistics in the overlapping regions of Figure~\ref{f:simulation_matching} match on a reasonable level in both overlap regions. Hence, it is acceptable to patch our three simulations together to obtain the stick distribution across all considered timescales.  We note that for stick timescales approaching 1 Gyr, the number of sticks in our simulations becomes small and thus susceptible to notable Poisson errors.  The additional variability and fall-off in cumulative resonance time near $10^9$ years in Figure~\ref{f:simulation_matching} are likely due to small-number statistics.   We also note that the largest stick-duration bin, which includes particles that remained in resonance at the very end of the simulation, does not contain a substantial excess of particles.

\section{Results}\label{s:results}

We find that a significant fraction of the scattering population is transiently stuck in resonances at any given time (Section \ref{sec:total}). After presenting our predictions for the relative population in each resonance, we discuss the distribution of transiently-stuck TNOs across different resonances (Section \ref{sec:pops}). We investigate the libration amplitude distributions for these objects in Section \ref{sec:amps}. In some cases, these population distributions depend on the duration of resonance sticks considered, and we comment on these differences in Section \ref{sec:difftimescales}.
Section \ref{sec:discussion} presents results specific to the 3:2, 2:1, and 5:2 resonances, where many observed objects are found.

\subsection{Total Population of Transiently-Stuck Resonant Objects}\label{sec:total}

Since our simulated scattering+transiently-stuck population is in pseudo-steady state, the time-weighted percentage of simulated objects with semi-major axes between 30 and 100~au that are in each dynamical class (i.e., in a resonance or non-resonant) corresponds to the fraction of the scattering+transiently stuck population that is currently in that dynamical class.  (Recall that our initial conditions are consistent with the orbital distribution of the currently observed non-resonant scattering population.)
 To calculate these time-weighted percentages, we sum the cumulative stick times in each resonance, weighted for each simulation and stick timescale as described in Section \ref{ss:datamp}, then divide by the total normalized time spent by simulated particles in the range 30-100au, $T_{range,15Myr}(1{\rm Gyr}/15 {\rm Myr})$.

\begin{figure*}[htbp]
   \centering
   \vspace{-0.5in}
   \includegraphics[width=4.5in]{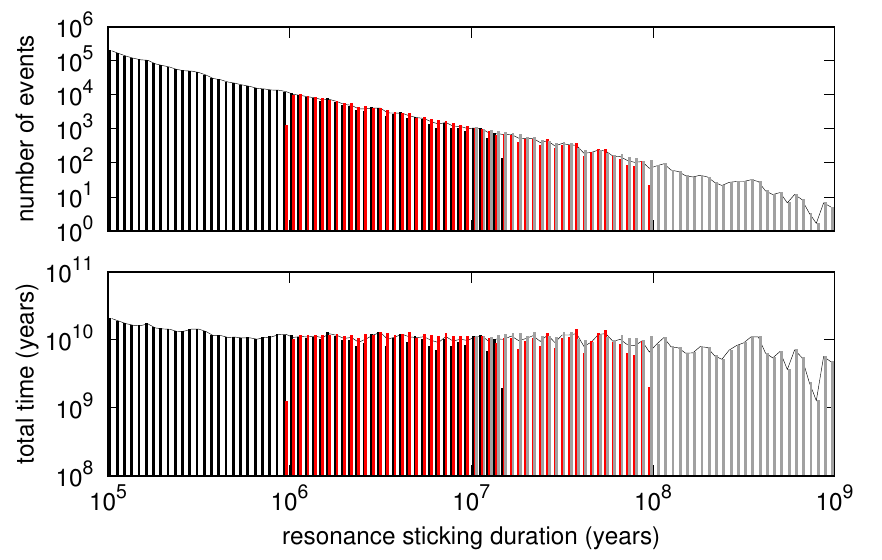}
   \caption{Normalized number of resonance sticks (top) and cumulative time spent in resonance (bottom) for all test particles as a function of the duration of a single stick. The cumulative resonance time is proportional to the likelihood of observing a resonant object at a snapshot in time (e.g., today) as a function of stick duration. Histogram counts are calculated for log bins in stick time spaced 0.04 apart. Results from Simulations 1 (black), 2 (red), and 3 (gray; see Table~\ref{tab-sticks} for simulation parameters) are shown separately as bars; each simulation is individually normalized using steps 1 and 2 as described in Section~\ref{ss:datamp}. The thin black line shows the result of combining all three simulations by following steps 1-3 in Section~\ref{ss:datamp}.}\label{f:simulation_matching}
\end{figure*}

Using this procedure, we find that the fraction of the scattering+transiently-stuck population that is currently in resonance is $\sim0.4$; the right-most column in Table~\ref{tab-res-frac} gives this fraction for the entire $a=30-100$~au population, as well as the fraction for each resonance identified in our simulations. 
Table~\ref{tab-res-frac} also lists the time-weighted resonant fractions for each of our three simulations (where each stick is time weighted, but there is no normalization to account for the different simulation lengths or the effects of population decay).
We note that the time-weighted resonant fraction for simulation 1 (our 15 Myr simulation) is 0.25, while the fraction is 0.17 for simulation 3 (our 1 Gyr simulation). These two simulations barely overlap in terms of the resonance stick resolution, which is why the combined, normalized time-weighted resonance fraction for all of our simulations is very roughly the sum of these two fractions.
A comparison of the three simulation (and stick resolution) timescales tells us that short sticks dominate the transient sticking population in number but only marginally in total time in resonance. This can also be seen from Figure~\ref{f:simulation_matching}.
Thus, perhaps counter-intuitively, we predict that the likelihood of observing a transiently-stuck object that is expected to stay in resonance for $\sim$1~Gyr is similar to the likelihood of observing an transiently-stuck object that is unstable on a short timescale.

\begin{rotatetable*}
\begin{deluxetable*}{c|ccc|ccc|ccc|c}
\tabletypesize{\footnotesize}
\tablecolumns{10}
\tablewidth{0pt}
\tablecaption{Time spent by simulated particles between $30<a<100$~au: total and by resonance\label{tab-res-frac}}
\tablehead{
\colhead{} & \multicolumn{3}{c}{Simulation 1 (15 Myr)} & \multicolumn{3}{c}{Simulation 2 (100 Myr)} &  \multicolumn{3}{c}{Simulation 3 (1 Gyr)} &\colhead{time-weighted} \\ 
\colhead{particles}& \colhead{ sticks } 	& \colhead{time (yr)}  & \colhead{frac} & \colhead{ sticks } 	& \colhead{time (yr)}  & \colhead{frac}& \colhead{ sticks } 	& \colhead{time (yr)}  & \colhead{frac} & \colhead{combined frac.}}
\startdata
all particles & -- & $3.86\times10^{10}$ & 1.0 & -- & $2.37\times10^{11}$ & 1.0 & -- & $1.53\times10^{12}$ & 1.0 & 1.0 \\
\hline 
\uline{resonance} & & & & & & & & & \\
2:1 & 169 & $1.02\times10^8$ &  $2.65\times10^{-3}$ & 110 & $4.35\times10^8$ & $1.84\times10^{-3}$ & 49 & $1.91\times10^{9\phn}$ & $1.25\times10^{-3}$ & $3.55\times10^{-3}$ \\
3:1 & 298 & $1.63\times10^{8}$ & $4.22\times10^{-3}$ & 288 & $1.51\times10^{9}$ & $6.37\times10^{-3}$ & 149 & $7.15\times10^{9\phn}$ & $4.66\times10^{-3}$ & $9.56\times10^{-3}$\\
4:1 & 365 &  $2.18\times10^{8}$ & $5.66\times10^{-3}$ & 356 & $2.14\times10^{9}$ & $9.03\times10^{-3}$ & 283 & $1.30\times10^{10}$ & $8.45\times10^{-3}$ & $1.44\times10^{-2}$\\
5:1 & 437 & $2.90\times10^{8}$ & $7.51\times10^{-3}$ & 424 & $2.51\times10^{9}$ & $1.06\times10^{-2}$ &  587 & $3.20\times10^{10}$ & $2.08\times10^{-2}$ & $2.60\times10^{-2}$\\
6:1 & 497 & $5.57\times10^{8}$ & $1.44\times10^{-2}$ & 762 & $6.32\times10^{9}$ & $2.67\times10^{-2}$ & 1004 & $4.69\times10^{10}$ & $3.06\times10^{-2}$ & $4.21\times10^{-2}$\\
3:2 & 51 & $2.77\times10^{7}$ & $7.16\times10^{-4}$ & 45 & $1.76\times10^{8}$ & $7.41\times10^{-4}$ & 9 & $1.49\times10^{8\phn}$ & $9.73\times10^{-5}$ & $9.30\times10^{-4}$\\
5:2 & 112 & $6.12\times10^{7}$ & $1.59\times10^{-3}$ & 57 & $2.20\times10^{8}$ & $9.28\times10^{-4}$ & 24 & $1.12\times10^{9\phn}$ & $7.30\times10^{-4}$ & $2.12\times10^{-3}$\\
\vdots & \vdots & \vdots & \vdots & \vdots & \vdots & \vdots & \vdots & \vdots & \vdots & \vdots\\ 
all & 24089 & $9.51\times10^{9}$ & 0.246 & 11348 & $4.86\times10^{10}$ & 0.205 & 7083 & $2.68\times10^{11}$ & 0.175 & 0.403\\[-6pt]
resonant & & & & & & & \\	 
\enddata
\tablecomments{The time-weighted combined fraction is calculated using the normalization procedure outlined in Section~\ref{ss:datamp}. Only notable resonances are shown here. The full table is available as a machine readable file.}
\end{deluxetable*}
\end{rotatetable*}

We note that more than half of the scattering disk test particles that visit the semimajor axis range 30-100~au experience at least one sticking event in our $1.5\times10^7$-year simulation and approaching half experience at least one longer-duration sticking event in our $10^9$-year simulation (see Table~\ref{tab-sticks}).  In fact, the scattering population and the transiently-stuck population should be considered a single dynamical population.  This means that their size distributions, colors, etc. should be the same and that we may be able to use the physical properties of the scattering disk to identify transiently-stuck objects within resonances.

\subsection{Relative Populations Across Resonances}\label{sec:pops}

\begin{figure}[htbp]
   \centering
   \includegraphics[width=3.25in]{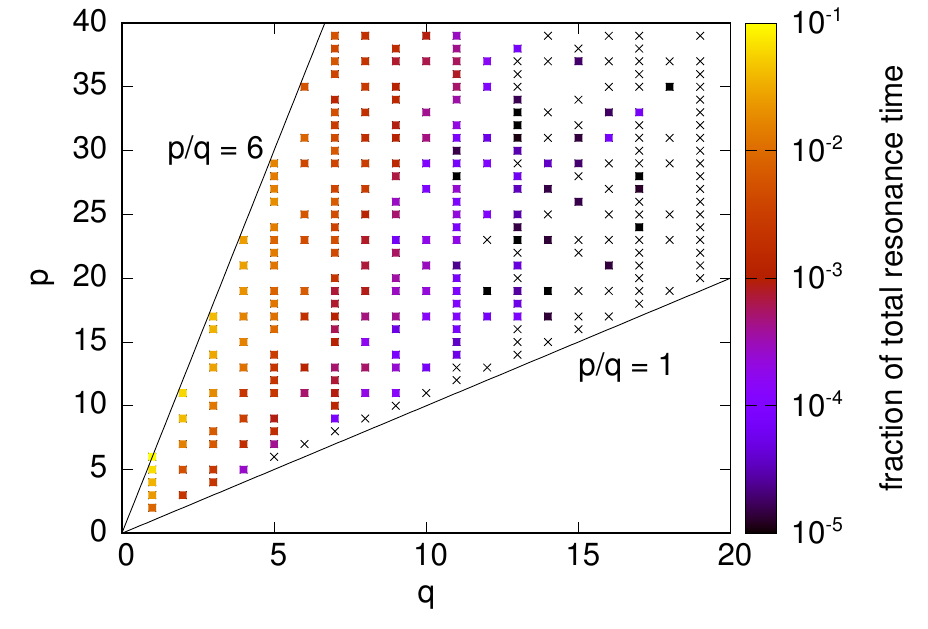}
   \caption{Map of the $p$:$q$ resonances checked by our resonance identification procedure (Section~\ref{ss:resid}). The colored boxes represent resonances with identified sticks, color coded by the fraction of the total normalized stick times spent in that resonance (sticks from the three simulations in Table \ref{tab-sticks} are combined using the normalization method described in Section \ref{ss:datamp}). Black crosses indicate resonances that were checked but had no sticks identified. Gaps occur where $p/q$ is the same as that of a lower-order resonance. The black solid lines indicate the period ratio range we considered.}\label{f:pqmap}
\end{figure}

\begin{figure*}[htbp]
   \centering
   \includegraphics[width=4.8in]{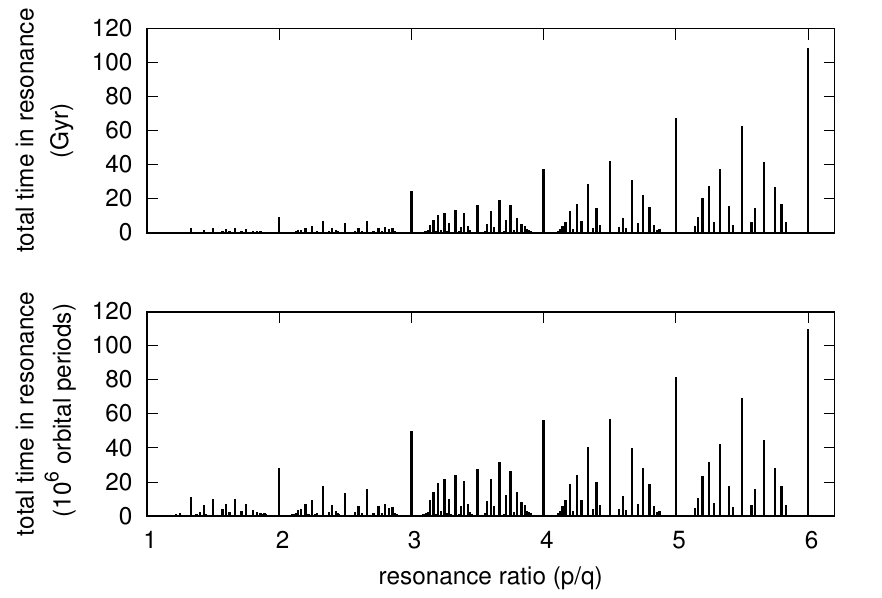}
   \caption{Total time spent by test particles in $p:q$ resonances in the range 30-100~au as a function of resonance ratio ($p/q$).  Sticks from the three simulations in Table \ref{tab-sticks} are combined using the normalization method described in Section \ref{ss:datamp}.  Total simulated particle time in Gyr (top) is proportional to the number of particles predicted to be each resonance at a snapshot in time (e.g., today).  Even when scaled to each resonance's orbital period (bottom), the total time spent by particles in resonance is generally larger for more distant resonances with the same value of $q$.  We suggest that the pattern evident in this figure results from total stick times proportional to the libration period in each resonance (c.f. Figure~\ref{f:theory_period2}).}\label{f:totalsticktimes}
\end{figure*}

Table~\ref{tab-res-frac} breaks down our prediction for the fraction of the scattering+transiently-stuck population into the fraction in each individual resonance.  Figures~\ref{f:pqmap} and \ref{f:totalsticktimes} show the total time spent by test particles sticking to resonances as a function of $p$ and $q$ separately and of the resonance ratio $p/q$, respectively.  This total time is summed across all particles in our three simulations using the normalization procedure described in Section \ref{ss:datamp}.  For any two  resonances, the ratio between the ``total sticking time" calculated for each population represents the relative likelihood of finding an object stuck to those two resonances at a snapshot in time (e.g., today).

The time-weighted fractions shown in Figure~\ref{f:pqmap} give us confidence that our resonance search (described in Section~\ref{ss:resid}) identified all of the most important $p$:$q$ resonances in the range $30<a\le100$~au. While there are an infinite number of possible $p$:$q$ pairs in the period ratio range $1/6 \le p/q < 1$, we imposed limits of $q<20$ and $p<40$. Figure ~\ref{f:pqmap} shows that the largest $q$ value for which we detected any resonance sticking was $q=18$, well within our imposed limit.  
It is also clear that the time-weighted fraction of the stuck population for this period ratio range decreases as $p$ approaches our limit of $p<40$; thus any potential $p$:$q$ resonances that fell outside our allowed range of $p$ are likely unimportant.

\begin{figure}[htbp]
\centering
	\includegraphics[width=3in]{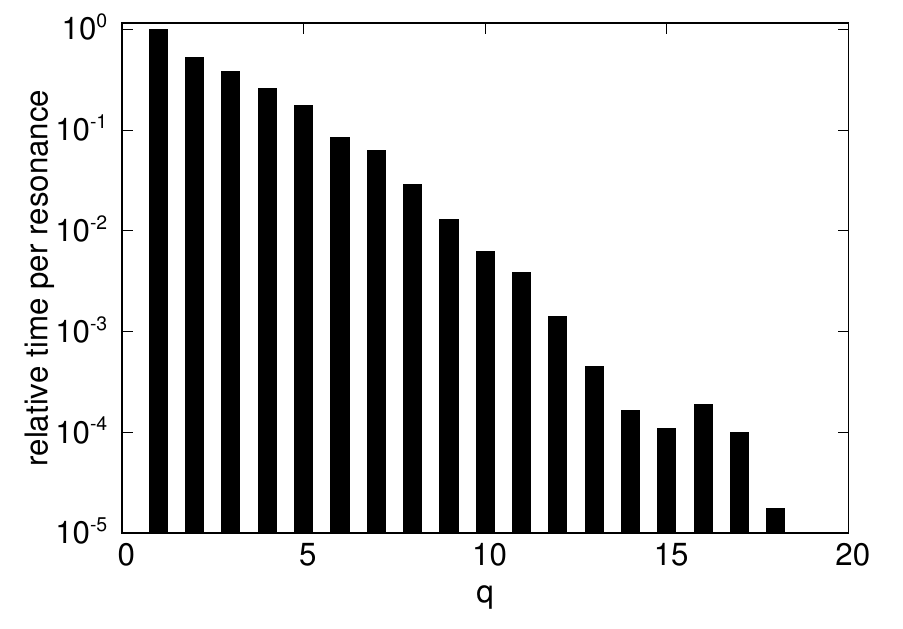}
	\caption{Normalized total duration of resonance sticking events per $p$:$q$ resonance as a function of $q$. Results from the three simulations in Table \ref{tab-sticks} are combined using the normalization described in Section \ref{ss:datamp}. For each value of $q$, the total time for all $p$:$q$ sticking events is divided by the the number of different possible values of $p$ in the range 30-100~au included in our resonance identification procedure (see Section~\ref{ss:resid}); the y-axis scale is normalized to the total time spent per $p$:1 resonance.}\label{f:average_duration}
\end{figure}

Structure is evident in the distribution displayed in Figure \ref{f:totalsticktimes}.  First, as reported in \citet{Lykawka2007}, resonances with smaller values of $q$ have larger predicted transiently-stuck populations; in particular, the $p$:1 resonances are the stickiest.  We display this result explicitly in Figure~\ref{f:average_duration}, which shows the normalized total particle time spent in resonance as a function of $q$.  For each value of $q$, we sum the normalized stick times (as described in Section \ref{ss:datamp}) for all $p$:$q$ resonances with identified sticks and then divide by the number of $p$:$q$ resonances included in our resonance identification procedure (described in Section~\ref{ss:resid}) to determine the average amount of time spent per resonance as a function of $q$; for example, we checked for five possible resonances in our $q=1$ bin (the 2:1 through 6:1 resonances), so the sum of the normalized stick times for that bin is divided by five.
Each bin is then normalized by the $q=1$ bin.
From Figure~\ref{f:average_duration} we see that resonance stickiness clearly decreases as $q$ increases (from $p$:1 to $p$:2 to $p$:3 and so on). This is similar to the trend found by \citet{Lykawka2007} (see their Figure 2).
Given that relative stickiness appears to be a function of $q$, all subsequent discussion will focus on using $q$ to characterize the transient population instead of the commonly-used resonance order ($p-q$).

Second, at fixed $q$, the total particle-time in resonance increases for larger $p/q$, which corresponds to larger orbital periods \citep[as also reported in][]{Lykawka2007}.  This trend is most obvious for the $p$:1 resonances. In Figure~\ref{f:totalsticktimes}, we show that this increase in stickiness with orbital period persists even when normalized to the local Keplerian orbital period for each resonance, $(p/q)P_{N}$, where $P_{N}$ is the orbital period of Neptune.

The combined dependence of total particle-time on $q$ and on orbital period produces a distinctive pattern in Figure \ref{f:totalsticktimes}. We suggest that this number comes from a combination of the distribution of semi-major axes for source particles in the scattering population and the resonance libration period in each resonance. This suggestion arises from the fact that the number of objects in resonance is proportional to the likelihood that an object sticks to the resonance (which is in turn proportional to the number of objects available for sticking) as well as to the time that each stuck object spends in resonance.

-The density of objects available for sticking comes from the scattering population from which transiently-stuck objects are drawn.  Because these particles experience a random walk in angular momentum with kicks occurring at a roughly fixed pericenter distance, the typical time for a scattering object to pass through semi-major axis $a$ is proportional to $v_{p}^2P$, where the pericenter velocity $v_{p} \propto a$ at large eccentricity and the orbital period $P \propto a^{3/2}$ (the factor $v_{p}$ comes from the random walk due to velocity kicks at pericenter, and the factor $P$ comes from the time between kicks).  Thus, the number of objects in the steady-state scattering population is roughly proportional to $a^{7/2}$.  We have confirmed that the model of the scattering population from which we draw our initial conditions satisfies this expression in the region $30<a<100$~au.

\begin{figure}[htbp]
   \includegraphics[width=3.3in]{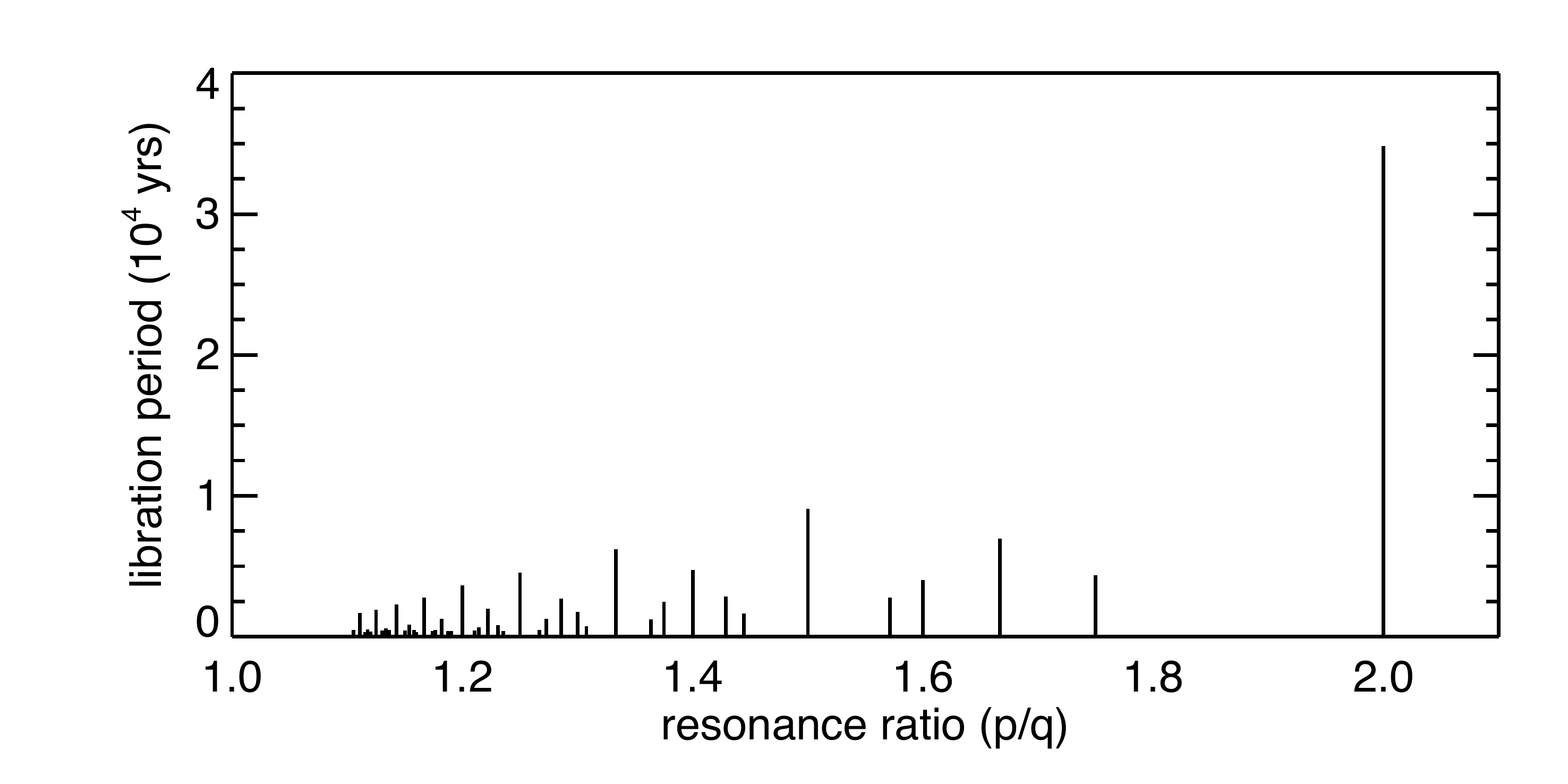}
   \caption{Analytic estimates of the resonant libration period as a function of resonance ratio, $p/q$, for resonances with order $p-q \le 4$.  Libration periods are calculated using expressions from \citet{MurrayDermott1999}. Note, there are many resonances for which we identified sticks that are not included in this plot due to the limited expansion of the disturbing function.}\label{f:theory_period2}
\end{figure}

Since the instantaneous stability to perturbations of a resonant orbit is typically a function of where the particle is in its libration cycle, it is reasonable to suggest that a typical timescale of a transient stick is measured in resonance libration periods.  
Figure \ref{f:theory_period2} displays the libration periods for resonances up to 4th order in the region 30-50au, estimated using the analytic expressions provided in \citet{MurrayDermott1999}.  The libration period is calculated using the the lowest-order resonance for each ratio $p/q$.  Because stuck objects are drawn from the scattered population, we evaluate these expressions using eccentricity $e = 1 - (q/p)^{2/3}$, which corresponds to an orbit with pericenter at the semi-major axis of Neptune. (Scattered objects typically have pericenter distances larger than $a_{N}$ by several au---our choice for $e$ maintains simplicity while capturing the salient features in Figure~\ref{f:theory_period2}.) Unfortunately, the structure in Figure~\ref{f:totalsticktimes} is clearest and least subject to small-number statistics for resonances of order $>$4, while the expansions provided in \citet{MurrayDermott1999} are available only up to 4th order, so we cannot directly confirm that total sticking times are proportional to libration period simply by normalizing our simulation results to the analytic estimates.  However, the pattern in Figure~\ref{f:theory_period2} matches that in Figure~\ref{f:totalsticktimes} qualitatively well. Libration periods are longer for resonances with lower values of $q$, producing the distinctive nested-triangle pattern present in both plots.

Putting these two considerations together, we interpret the structure in Figure \ref{f:totalsticktimes} as proportional to the number of source objects at each semi-major axis (set by the scattering population) and to the resonance libration period in each resonance (due to the stick timescale).  This interpretation is consistent with the fact that a plot (not shown) analogous to Figure \ref{f:totalsticktimes} but showing stick number rather than total stick times yields a much flatter distribution for adjacent resonances and has an envelope well-fit by the function $a^{7/2}$.  More detailed modeling may yield additional, more subtle, dependencies due to variations in the source population density in phase space compared to the phase space volume of each resonance.

\subsection{Libration Amplitude Distributions}\label{sec:amps}

\begin{figure*}[htbp]
	\centering
	\includegraphics[width=6.5in]{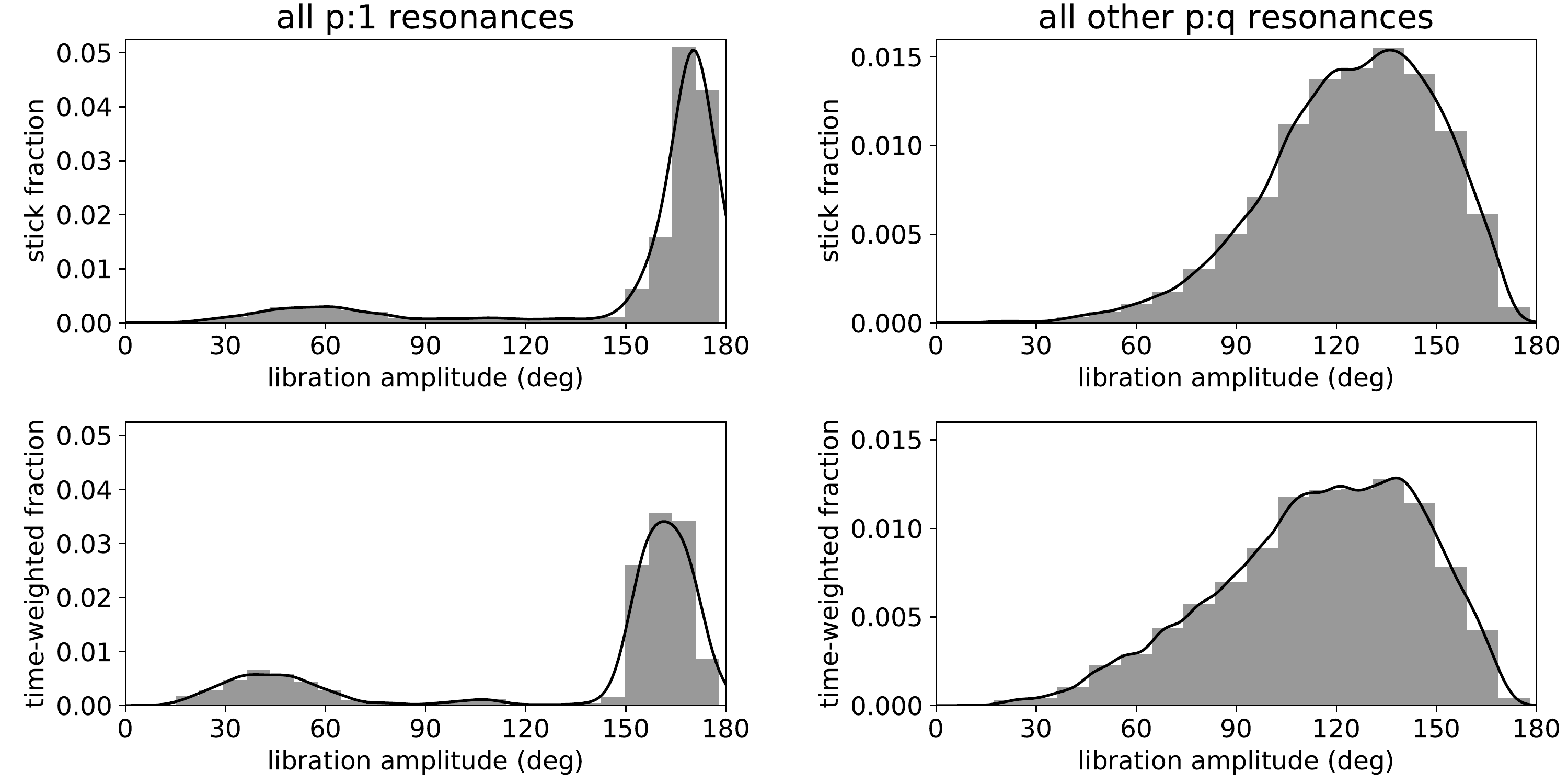}
	\caption{Distribution of the number of sticking events in the simulations as a function of libration amplitude (top panels) and the time-weighted fraction of sticking events as a function of libration amplitude (bottom panels). We separated $p$:1 resonances (left panels) from the rest of the resonances (right panels) to show the bimodal nature of the $p$:1 libration amplitude distribution due to the existence of both symmetric and asymmetric libration islands. The gray histograms are the binned results of all three simulations (combined using the normalization method described in Section~\ref{ss:datamp}), and the solid black lines are kernel density estimates for the same data. Histograms are normalized such that the area under each curve adds to 1. }\label{f:amp_time_count_all}
\end{figure*}

Next, we examine the libration amplitude distribution for transiently-stuck particles.  Figure \ref{f:amp_time_count_all} displays the fractional number of sticking events (top panels) and the time-weighted fraction of events (bottom panels) as a function of libration amplitude.  Data from the three simulations in Table \ref{tab-sticks} are combined using the normalization method in Section \ref{ss:datamp}. To build up statistics, we combine events from multiple resonances.  The solid black lines overplotted on the histograms provide kernel density estimates to mitigate the impact of our choice of bin size.  Because $p$:1 resonances exhibit asymmetric libration, we separate $p$:1 resonances (left panels) from all other resonances (right panels).  The possibility of asymmetric libration produces two peaks in the $p$:1 distribution.  We note that the small number of sticks midway between the two peaks is likely an artifact of our averaging method for computing the libration period, resulting from sticks that jump between symmetric and asymmetric libration rather than from sticks that consistently exhibit intermediate libration periods.

\begin{figure}[htbp]
\centering
	\includegraphics[width=3.25in]{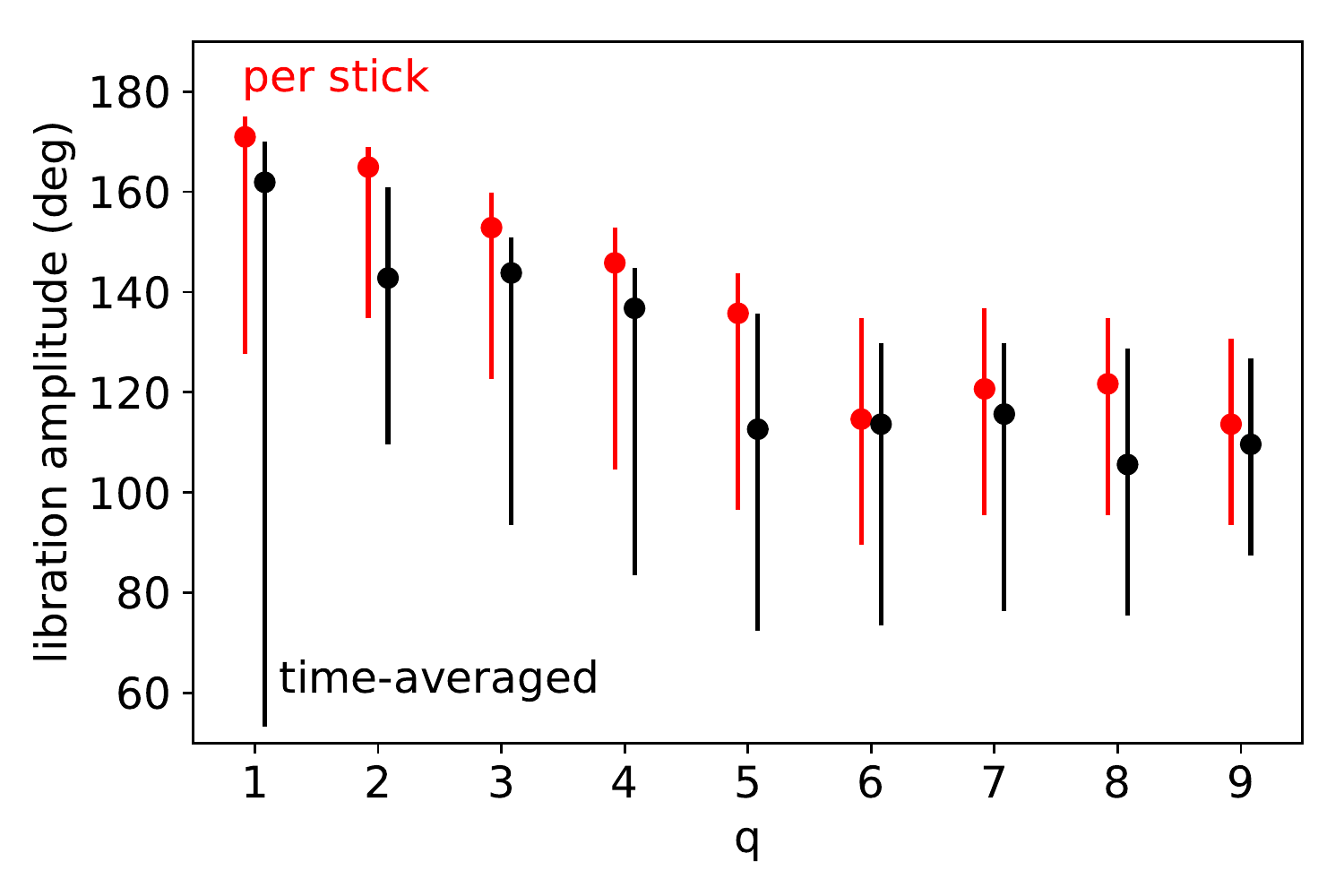}
\caption{The most probable amplitude (dots) and middle 1-$\sigma$ portion of the amplitude distribution for individual sticking events by number (red) and for the time-weighted resonance sticking population (black) in p:q resonances from 30-100 au as a function of q. We plot only values of q for which at least $\sim1000$ sticks were recorded. The very wide range of amplitudes for the time-averaged $p$:1 population is due to the longer-lived nature of the low-amplitude sticks in the asymmetric $p$:1 islands (see Figure~\ref{f:amp_time_count_all}). The sticking amplitude appears to be be inversely related to q.}\label{f:time_amp_prob}
\end{figure}

Resonances with $q>1$, in contrast exhibit a single broad peak in libration amplitude, strongly skewed toward values in excess of 90 degrees.  The breadth of the peak in Figure \ref{f:amp_time_count_all} comes in part from the wide range of resonances included in the average.  As shown in Figure \ref{f:time_amp_prob}, libration amplitudes are generally smaller for sticks with larger values of $q$.  Figure \ref{f:time_amp_prob} displays this dependence on $q$ in two ways: we plot the most probable amplitude (analogous to the peak in the kernel density estimates in Figure \ref{f:amp_time_count_all}) as dots and the middle 68.3\% (1-$\sigma$) of the libration amplitudes as vertical bars.  For resonances with $q < 5$, smaller $q$ corresponds to larger typical libration amplitude.

Figure \ref{f:time_amp_prob} provides typical libration amplitudes both per stick (red) and for our time-averaged population (i.e., by total stick time in the simulation; black).  The time-averaged results skew to lower libration amplitudes because, in general, sticks with lower libration amplitudes are longer-lived (we have verified this directly by plotting average stick times as a function of libration amplitude; not shown).  This effect is particularly pronounced for the $p$:1 resonances, for which objects in the asymmetric islands, which are longer-lived than symmetric librators, have much smaller libration amplitudes.  In Section \ref{sec:discussion}, we provide libration amplitude distributions for a few resonances of particular observational interest.

\clearpage

\subsection{Differences Between the Three Simulation Timescales}\label{sec:difftimescales}

\begin{figure}[htbp]
\centering
	\includegraphics[width=3.25in]{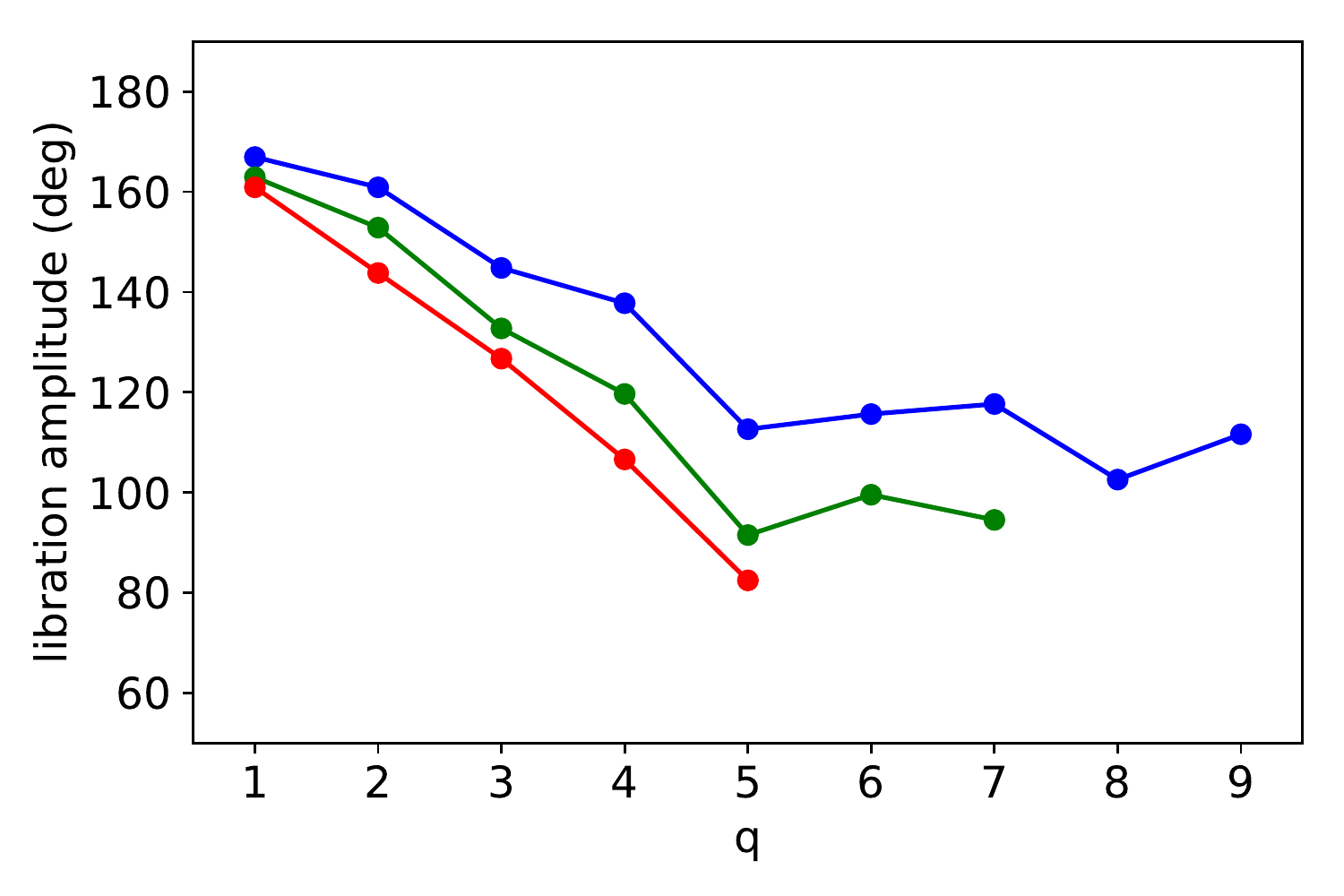}
	\caption{The time-weighted most probable amplitude of resonance sticking in p:q resonances from 30-100au as a function of q (compare with the black dots in Figure \ref{f:time_amp_prob}). From top to bottom (blue, green and red), the lines represent data from the 15~Myr, 100~Myr, and 1Gyr simulations. For each simulation we include only values of q for which $\sim1000$ sticks were recorded. At each q, amplitudes are smaller for longer simulations (which measure longer stick timescales).}\label{f:time_amp_prob_ind}
\end{figure}

We have explained that the difference between the number-weighted and time-weighted amplitude distributions in Figure~\ref{f:time_amp_prob} results from an inverse correlation between the amplitude of libration for a resonance stick and the stick's duration. If so, the most probable amplitude should be different for our different simulation time scales.  We verify this difference in Figure \ref{f:time_amp_prob_ind}, which is analogous to Figure \ref{f:time_amp_prob} but displays only the time-averaged most probable amplitude for clarity.  Sticks that occur in the longest simulation (which have on average the longest sticking timescales) typically occur at lower libration amplitudes.

\begin{figure}[htbp]
 	\centering
\includegraphics[width=3.25in]{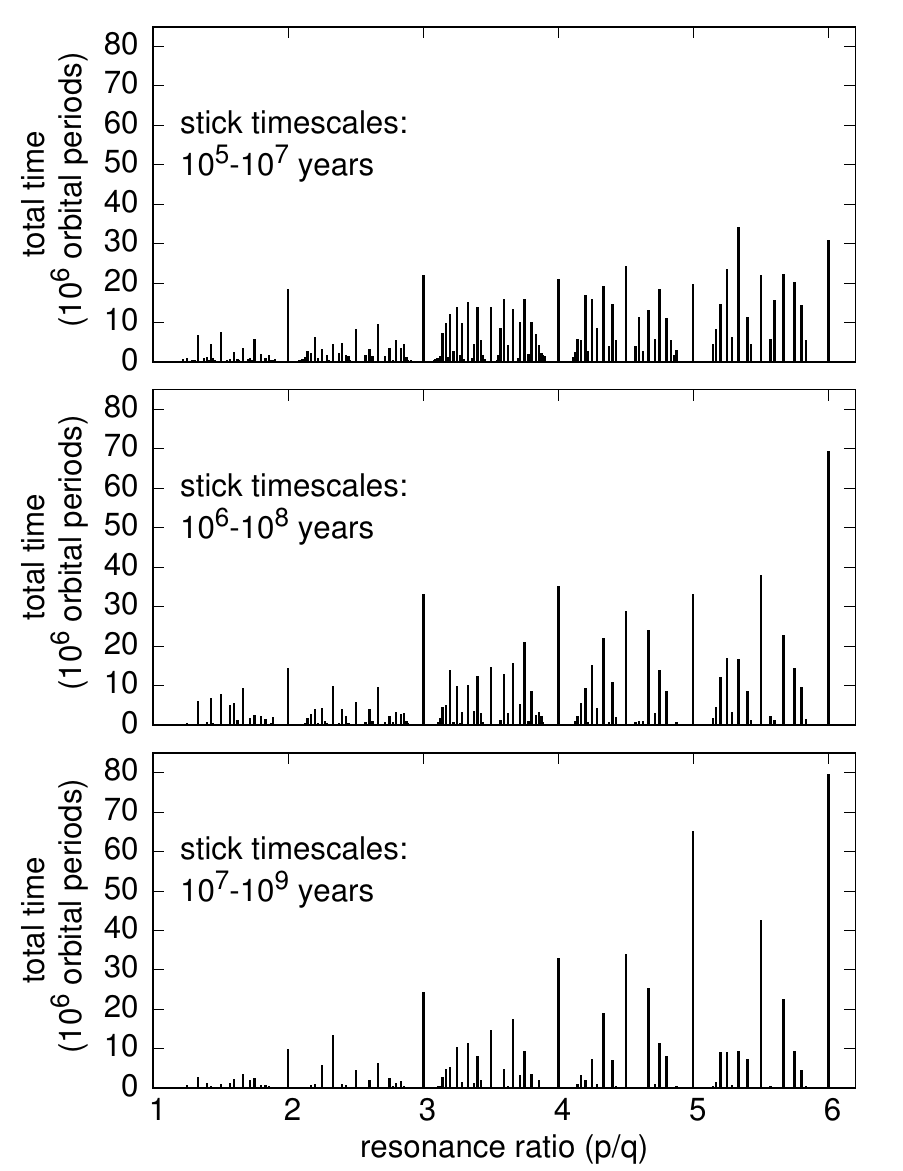}
\caption{The same as Figure \ref{f:totalsticktimes}, separated into results for the 15Myr (top), 100Myr (middle), and 1Gyr (bottom) simulations.  Stick timescales greater than 10Myr are omitted from the top panel so that each panel shows the same number of log bins in stick timescale.  Normalization is as in Figure \ref{f:totalsticktimes} except that overlapping stick timescales are counted fully in each plot.}\label{f:orbital_three_simulation}
\end{figure}

This difference in stick duration and hence typical libration amplitude generates subtle differences in the distribution across resonances for our three simulation timescales.  Figure \ref{f:orbital_three_simulation} provides the same information displayed in Figure \ref{f:totalsticktimes}, but separated into results for the simulations in Table \ref{tab-sticks}.  The same normalization is used as for Figure \ref{f:totalsticktimes}, except that the overlapping regions are counted fully in each panel, and we omit sticks lasting more than $10^7$ years from Simulation 1 so that each panel covers the same number of log bins in stick timescale.

Figure \ref{f:orbital_three_simulation} demonstrates that $p$:1 resonances are most prominent for long stick timescales.  In contrast, short stick timescales dominate high-$q$ resonances.  The structure between each $p$:1 resonance becomes less and less dense---hosting fewer sticks with large $q$ values---as the measured stick timescale increases.

We note that changing the maximum libration amplitude for which we will consider test particles in the simulations to be resonant from our fiducial $175^{\circ}$ to a smaller value can impact the inferred relative resonance populations. This is particularly true for the $p$:1 resonant populations. From Figure~\ref{f:amp_time_count_all}, it's clear that a maximum libration amplitude cut even as low as $150-160^\circ$ would only minimally impact the number of sticks detected for the non $p$:1 resonances; a cut at $160^\circ$ would eliminate only $\sim$10\% of all sticks and $\sim$5\% of the time-weighted population in these resonances. In contrast, such a cut would eliminate $\sim$70\% of the $p$:1 sticks and $\sim$50\% of the time-weighted $p$:1 populations. Even shifting the maximum libration amplitude from $175^\circ$ to $170^\circ$ would reduce the time-weighted $p$:1 population by $\sim10$\%. This has important implications for the comparison of our estimated stuck populations to observational constraints; if the maximum libration cut used to dynamically classify the observed objects is very different from our cut, then the predicted stuck population should be appropriately adjusted. Another factor to consider is that real objects have observational uncertainties associated with their observed orbits and thus with their dynamical classifications. Real objects with large libration amplitudes that are currently near the edges of resonances might be mis-classified as non-resonant if their orbits are not well enough determined.

\section{Discussion and Comparison to Observations}\label{sec:discussion}

For use in comparison with observations, we provide explicit predictions in Section~\ref{sec:predict} for the number of objects with $H_r < 8.66$ currently transiently-stuck in all of the resonances identified in our simulations, and we compare these to observationally derived population estimates in the literature. In Section~\ref{sec:specific}, we also discuss the libration amplitude distributions for objects transiently stuck in the 3:2, 2:1, and 5:2 resonances.  The observational constraints discussed below indicate that these resonances are not dominated by transiently stuck objects (consistent with theoretical expectations for the 3:2 and 2:1 resonances), but we discuss how the libration amplitudes of transiently stuck objects might be used to help identify transient interlopers in resonances dominated by presumably primordial populations.

\subsection{Predictions for the transient resonance populations}\label{sec:predict}

In Table~\ref{tab-res-preds}, we translate the fractional resonance populations from Table~\ref{tab-res-frac} into predictions for the number of transiently-stuck objects with $H_r < 8.66$ currently in each resonance.  To generate these predictions, we use the constraints on the scattering population from the Outer Solar System Origins Survey \citep{Bannister2016,Bannister2018}. Based on the number scattering objects detected in this survey, \citet{Lawler2018} estimate that there are $1.1\pm0.2\times10^4$ scattering objects with $H_r < 8.66$ and semimajor axes in the range 30-100~au. Producing this estimate relies on assuming a model for the true orbital and H magnitude distribution of the scattering population. \citet{Lawler2018} uses the same \citet{Kaib2011} orbital model, modified as in \citet{Shankman2013} to account for the larger orbital inclinations in the observed population compared to the model. This is the same orbital distribution we take as our simulation initial conditions (Section~\ref{ss:sims}), so the \citet{Lawler2018} scattering population estimate will produce self-consistent predictions for our resonant populations. The population estimate above is for a divot model of the H magnitude distribution, but because we use an H magnitude cut $H_r < 8.66$, the scattering population estimate is not very sensitive to this choice of a divot instead of a knee in the H magnitude distribution (see Table 1 in \citealt{Lawler2018}, where the difference in the total scattering population with $H_r < 8.66$ for a divot vs a knee is only $\sim10$\%, smaller than the population uncertainty). 

To use this scattering population estimate for our predictions, we must also translate our fractional resonance populations from Table~\ref{tab-res-frac} from fractions relative to the scattering+stuck population to those relative to just the scattering population. The entire transiently stuck resonant population represents $\sim40\%$ of the scattering+stuck population, which translates to the stuck population being $\sim68\%$ of the number of scattering objects. For our population above, this means we expect $\sim7400$ transiently stuck resonant objects with $H_r < 8.66$ from $a=30-100$~au. For each individual resonance, we then calculate its fractional contribution to the overall resonant population to produce the individual population estimates in Table~\ref{tab-res-preds}.

\begin{deluxetable*}{cccccr}
\tabletypesize{\footnotesize}
\tablecolumns{6}
\tablewidth{0pt}
\tablecaption{Summary of normalized transient resonance fractions and predicted absolute transient sticking populations for the range $30<a<100$~au \label{tab-res-preds}}
\tablehead{
\colhead{resonance} & \colhead{semimajor} &\colhead{fraction of scattering+} & \colhead{fraction of stuck}& \colhead{fraction of scattering } & \colhead{approximate} \\[-8pt] 
\colhead{} & \colhead{axis} &\colhead{stuck population} &\colhead{population} & \colhead{population} & \colhead{predicted number} \\[-8pt] 
\colhead{} & \colhead{$a$ (au)} & \colhead{ $f_{s+r}$} & \colhead{ $f_{r}$} &\colhead{$f_s$} & \colhead{with $H_r<8.66$}
}
\startdata
all & 30--100 & 0.403 & 1.0 & $0.676$ & 7434 \\ \hline
2:1 &47.7 & $3.55\times10^{-3}$ & $8.81\times10^{-3}$ & $5.95\times10^{-3}$ &65\\
3:1 & 62.5 & $9.56\times10^{-3}$ & $2.37\times10^{-2}$ & $1.60\times10^{-2}$ &176 \\
4:1 & 75.8 & $1.44\times10^{-2}$ & $3.57\times10^{-2}$ & $2.42\times10^{-2}$ & 266 \\
5:1 &87.9 & $2.60\times10^{-2}$ & $6.46\times10^{-2}$ &  $4.36\times10^{-2}$ & 480 \\
6:1 & 99.2 & $4.21\times10^{-2}$ & $1.04\times10^{-1}$ & $7.06\times10^{-2}$ & 777 \\
3:2 & 39.4& $9.30\times10^{-4}$ & $2.31\times10^{-3}$ & $1.56\times10^{-3}$ & 17\\
5:2 &55.4 & $2.12\times10^{-3}$ & $5.27\times10^{-3}$ & $3.56\times10^{-3}$ & 39\\
\vdots & \vdots & \vdots &\vdots &\vdots &\vdots \\
\enddata
\tablecomments{For all the resonances combined, the stuck population represents $f_{s+r}/(1-f_{s+r})=0.676$ of the scattering population in the range $30<a<100$~au. For each individual resonance, the stuck population in that resonance then represents $f_r \times 0.676$ of the scattering population. We have translated the predicted fraction of resonant objects to an actual number of objects based on the \citet{Lawler2018} estimate that there are  $(1.1\pm0.2)\times10^4$ actively scattering objects with $30<1<100$~au based on the results of the Outer Solar System Origins Survey ~\citep{Bannister2016,Bannister2018}
; all predictions should be assumed to have an uncertainty of $\sim20$\% due to the observational uncertainty on the scattering population estimate. We list only resonances of note here; the entire table is available as a machine readable file. }
\end{deluxetable*}

The predicted stuck populations for the 3:2, 2:1, and 5:2 resonances are very low, of order a few tens of objects per resonance. For comparison, \citet{Volk2016} estimate that there are $8000^{+4700}_{-4000}$ 3:2 objects, $5700^{+7300}_{-4000}$ 5:2 objects, and $5200^{+900}_{-4000}$ 2:1 objects with $H_r~<~8.66$ based on an analysis of the detections in the first quarter of the Outer Solar System Origins Survey \citep{Bannister2016}. Earlier population estimates for these resonances based on the Deep Ecliptic Survey are slightly smaller \citep{Adams2014}, although still significantly larger than our predictions for the stuck population. It is thus clear that transient sticking from the actively scattering population is not a viable explanation for the unexpectedly large population of 5:2 objects.

Population estimates are available in the literature for several other resonances that may host significant transient sticking populations. \citet{Alexandersen2016} provided lower limits on the 3:1 and 4:1 populations based on observations, finding that they contain $>1100$ and $>24$ objects with $H_r~<~8.66$, respectively.  \citet{Pike2015} find $>10^4$ objects in the 5:1, when extrapolated to the same magnitude limit using the method described in \cite{Volk:2018}.  \citet{Volk:2018} estimate a population of $>4000$ objects in the 9:1 with $H_r~<~8.66$  based on two detections.  
These estimates exceed our population predictions for both the 3:1 and 5:1 resonances by factors that, though less than that for the 5:2, are quite large.  We note that the disagreement between our predicted 5:1 population and the observational population estimate is particularly interesting given that \citet{Pike2015} report that the four detected 5:1 objects appear to be unstable, leaving the resonance on $\sim10^7$ year timescales.

However, comparing our predictions to observations on an individual resonance-by-resonance basis may lead to spurious results---those resonances that happen to have detections will be interpreted as having large populations.  A more robust comparison for resonances with small numbers of observed objects would be to model the scattering+transient sticking population as a whole, which requires careful analysis of a well-characterized survey like the the Outer Solar System Origins Survey \citep{Bannister2018}.  We leave this---and the question of whether many distant resonances contain large non-transient populations---for future work.

\subsection{The 3:2, 2:1, and 5:2 resonances}\label{sec:specific}

Because the 3:2, 2:1, and 5:2 resonances currently host the most observed TNOs and are hence the most promising resonances within which to measure the libration amplitude distribution observationally, we provide a separate discussion for these resonances here. Even in resonances for which other dynamical mechanisms likely produced the majority of the resonance population, transient sticking nevertheless occurs.  Transiently-stuck interlopers should have a distribution of physical properties (such as photometric colors and binary fraction) that matches the scattering population.  They can also be dynamically distinguished by their distribution of libration amplitudes.  In practice, a combination of libration amplitudes and physical properties will be most convincing.

In order to accumulate sufficient statistics to plot the libration amplitude distributions for these resonances individually, we perform a fourth simulation.  We use the same initial conditions and simulation setup used for all simulations in Table~\ref{tab-sticks}, but run for a full 100Myr with the short output cadence used in Simulation 1.  In order to make the amount of output data manageable, we limit this simulation to output in the range 30-60~au, which covers the three resonances of interest. During the 100~Myr of this simulation, a total of 11617 transient sticking events were observed: 566 of them are 5:2 resonances, 355 of them are 3:2 resonances, and 829 of them are 2:1 resonances.  To account for long-duration sticks, we combine our results with sticks from Simulation 3 in Table~\ref{tab-sticks} in a manner analogous to that described in Section~\ref{ss:datamp}.

\begin{figure}[htbp]
\centering
\includegraphics[width=3.25in]{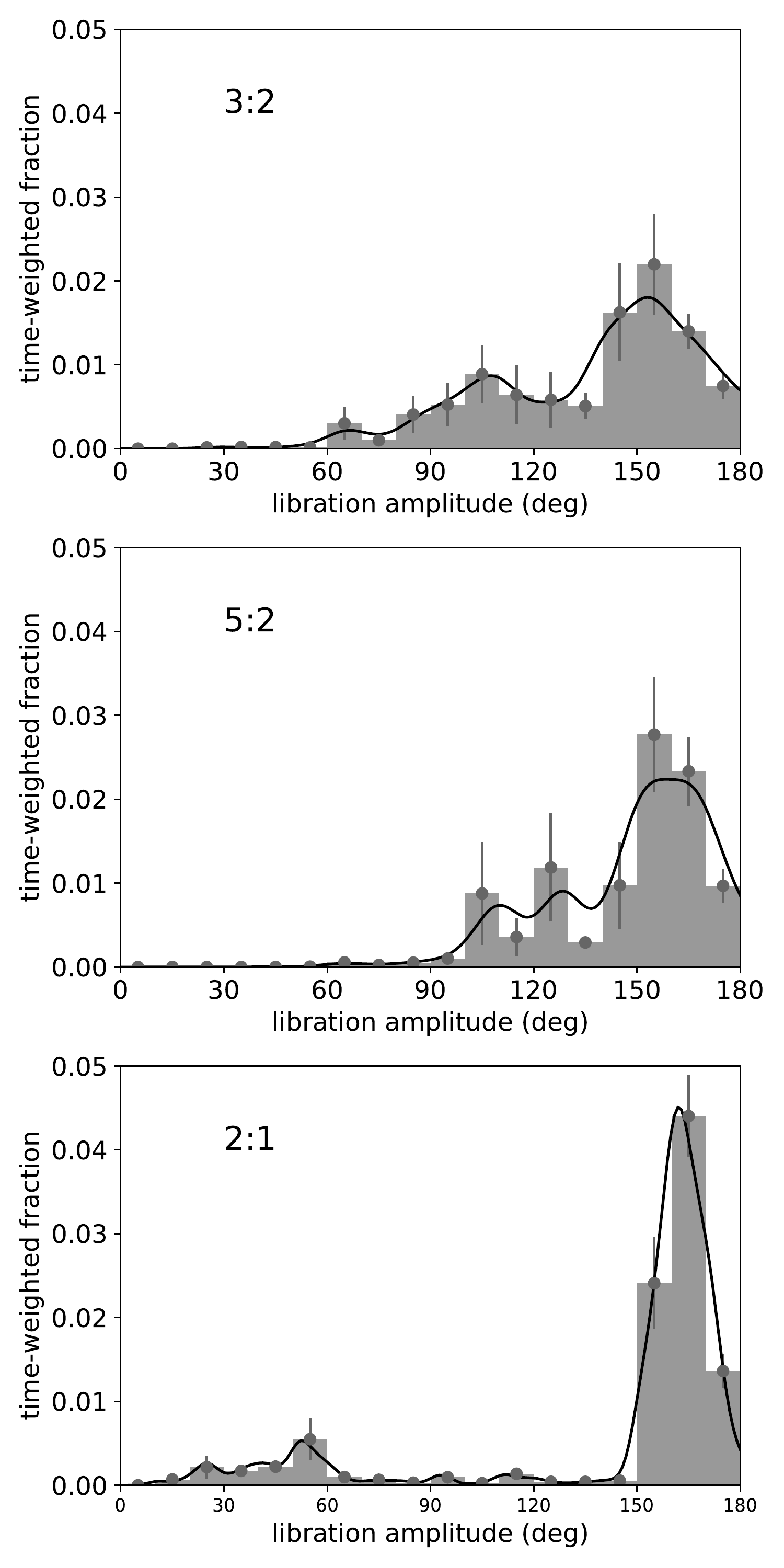}
	\caption{The normalized, time-weighted amplitude distributions for the 3:2, 5:2, and 2:1 resonances in the high resolution 100~Myr simulation combined with the 1~Gyr simulation.}
    \label{f:three_resonances_fourth_simulation}
\end{figure}

In Figure ~\ref{f:three_resonances_fourth_simulation}, we show the resulting time-weighted distribution of libration amplitudes for the 3:2 (top), 2:1 (middle), and 5:2 (bottom) resonances.  As in Figure~\ref{f:amp_time_count_all}, the solid lines provide kernel density estimates for the same data displayed in the histograms.  Because the number of sticks included in these plots are low enough that small number statistics may notably affect our results, we plot error bars for each histogram bin.  We bootstrap the error bars by drawing a total number of particles equal to that in the simulation from all simulation particles, with replacement, and recomputing the histogram.  We repeat this procedure 1000 times and plot the \textit{rms} variation of the resulting histogram heights as our error range in Figure~\ref{f:three_resonances_fourth_simulation}

It is clear that all three resonances have their most probable amplitude around or in excess of $150^\circ$. Comparing with Figure~\ref{f:amp_time_count_all}, the 2:1 distribution is similar to the overall $p$:1 distribution, with a slightly higher representation of symmetric librators.  The 3:2 and 5:2 distributions, however, show features that are smoothed over in the bottom-right panel of Figure~\ref{f:amp_time_count_all}.  The 3:2 distribution exhibits two peaks, perhaps due to variations resulting from some objects simultaneously occupying a Kozai resonance within the 3:2.  

The 5:2 plot also exhibits a wide knee at amplitudes substantially smaller than its peak.  The wiggles in this knee are the result of small number statistics---they arise from a few very long sticks.  When restricted to stick timescales less than 10Myr (not shown), the 5:2 libration amplitude distribution shows a single, well-behaved knee.  We nevertheless include the long timescale sticks in our plot because the lower-libration-amplitude component of the distribution is underestimated when they are omitted. Compared to the 5:2, the 3:2 resonance has a tail of low-amplitude librators extending to smaller amplitudes, and the fraction of objects with libration amplitudes less than $150^\circ$ is larger. 

\begin{figure}[htpb]
\centering
\includegraphics[width=3.25in]{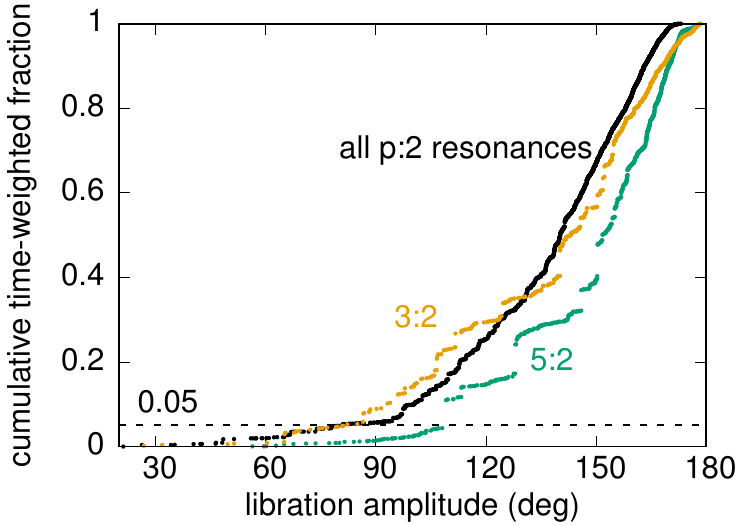}
\caption{The cumulative time-weighted libration amplitude distributions for all the $p$:2 resonances from the three simulations (normalized) and for the 5:2 and 3:2 resonances from the high-resolution 100 Myr simulation combined with the 1 Gyr simulation.}
\label{f:cumulative-amps}
\end{figure}

Figure \ref{f:cumulative-amps} provides the time-weighted, cumulative  libration amplitude distribution for all $p$:2 resonances in the three simulations in Table~\ref{tab-sticks} compared to the cumulative plot of the 5:2 and 3:2 libration amplitude distributions from Figure~\ref{f:three_resonances_fourth_simulation}. The libration amplitude distribution for all of $p$:2 resonances combined contains very few objects with amplitudes below 90$^\circ$, smoothly increases from $90-140^\circ$, has a broad peak, and then drops off to $175^\circ$.

We comment that \cite{Bannister201692} report the discovery of one 9:2 object, and their integrations of its observationally-determined orbit indicate that it will leave the 9:2 on $\sim$100~Myr timescales.  Its libration amplitude of 110$^\circ$ is consistent with our transient sticking results.

\subsection{Comparison with Observations in the 5:2}\label{sec:52}

The sky locations accessible to an object in resonance depends on its libration amplitude. Hence, a robust comparison between our data and the observed libration amplitude distribution in a resonance requires a well-characterized survey for which biases induced by pointing choices can be modeled.  The full dataset from the Outer Solar System Origins Survey \citep{Bannister2018} provides an ideal opportunity for such a comparison, and we leave this exercise for future work. 

Even without a full analysis, however, we can demonstrate that objects discovered during the first quarter of this survey do not appear consistent with our model of the current transiently-stuck population.  Four 5:2 objects were discovered during the first quarter of the Outer Solar System Origins Survey, with libration amplitudes of 62, 81, 88, and 122 degrees \citep{Bannister2016, Volk2016}.  Figure~\ref{f:cumulative-amps} displays our simulated cumulative time-weighted libration amplitude distributions for the 5:2.  We find that 95\% of the time-weighted 5:2 amplitudes are above 108 degrees, in strong disagreement with the low-amplitude librators (and absence of high-amplitude librators) reported. The Canada France Ecliptic Plane Survey \citep{Petit2011} also reported libration amplitudes for their five 5:2 detections; these best-fit amplitudes were all below $100^\circ$ \citep{Gladman2012},  further strengthening the case that stuck objects are not a significant contribution to the observed 5:2 population. (The cumulative distribution for all $p$:2 resonances is provided in Figure~\ref{f:cumulative-amps} for reference: 95\% of the time-weighted $p$:2 amplitudes are above 80 degrees.) 
We note that while there is a mild trend toward lower libration amplitudes at longer stick times, it is not strong enough to shift the amplitude distributions low enough to match the observational constraints when extrapolated to the full 4 Gyr lifetime of the solar system.
We conclude that it is unlikely to have more than a few objects in 5:2 with  libration amplitudes as low as indicated by the observations unless the objects have been stuck for considerably longer than 1~Gyr.

\section{Summary}\label{s:summary}

We simulate the population of TNOs drawn from the current scattering population that are transiently stuck into mean motion resonances with Neptune.  At a given snapshot in time for this pseudo-steady-state population in the region $30<a<100$~au, we find that about 40\% of all scattering+transiently stuck objects are transiently stuck in mean motion resonance, suggesting that the scattered disk and transiently-stuck resonant objects are best considered a single population (with the same distribution of physical characteristics). 

We measure the fraction of the transient+sticking population in each resonance with $1/6~\le~q/p < 1$, $p < 40$, and $q < 20$.  We confirm several results reported in \citet{Lykawka2007}:  the importance of transient sticking in $p$:$q$ resonances decreases with increasing $q$ (i.e. objects preferentially stick to p:1 resonances, then p:2 resonances, etc.); the number of objects stuck at a given time is larger for resonances at larger semimajor axes; and the transiently-stuck populations are dominated by large libration amplitudes. We additionally find that the transiently-stuck population of a resonance scales with both the population of scattering objects from which sticks are drawn and with the resonant libration period.  The number of objects increases with semi-major axis, set by the orbital random walk of the scattering population.  On average, stick timescales appear to scale approximately with libration period, generating the strong dependence on $q$.  This behavior likely results from some parts of the libration cycle having weaker stability to perturbations than others.

We translate our results into predictions of the total transient-sticking population with $H_r < 8.66$ based on a population estimate for the actively scattering population from \citet{Lawler2018}. We emphasize that these results refer only to sticks from the {\it current} scattering population and do not account for potential long-lived survivors of transient sticking during the early lifetime of the solar system, when the properties of the scattering population may have been different.  We predict that there should be $\sim7400$ transiently resonant objects in this magnitude range from $30<a<100$~au. Table~\ref{tab-res-preds} lists our predictions for all of the individual resonances identified in our simulations.
The number of objects in the 3:2, 2:1, and 5:2 resonances  inferred from observational constraints is larger than our transient sticking prediction, indicating that the majority of the objects in these resonances were likely emplaced in a different way.  For the 3:2 and 2:1 resonances, this is not surprising since standard dynamical models of the outer solar system emplace large populations of objects by other dynamical processes \citep[e.g.,][]{Hahn2005,Levison2008}.  For the 5:2, however, these models do not predict a large population,  suggesting that {\bf current} transient sticking is not a viable explanation for its large observed population.

We examined the libration amplitude distribution for our simulated stuck population, because libration amplitudes might provide another way to distinguish stuck objects from those emplaced by other dynamical mechanisms.
Considering a snapshot in time for $p$:1 resonances, we find that most stuck objects are symmetric librators with libration amplitudes larger than $150^\circ$;  $\sim22\%$ of the predicted $p$:1 population 
are asymmetric librators with amplitudes less than $90^\circ$.  When averaged together, $q \neq 1$ resonances also prefer high-amplitude librators, but the preference is not as severe. The peak in the libration amplitude distribution ranges from about 120-160$^\circ$, and the range of amplitudes encompassing the middle 68.3\% is 82-146$^\circ$.  For lower values of $q$, these distributions skew to higher amplitudes (c.f. Figures \ref{f:amp_time_count_all} and \ref{f:time_amp_prob}).
The libration amplitude distributions provide an additional check of the mismatch between the observed 5:2 population and our transient sticking model.
Specifically, we find that 95\% of the amplitudes of transiently-stuck 5:2 objects exceed 108$^\circ$.  Since three of the four 5:2 objects reported from the first quarter of the Outer Solar System Origins Survey  and all five reported 5:2 objects from the Canada France Ecliptic Plane Survey have libration amplitudes lower than this value \citep{Gladman2012,Volk2016}, the distribution of observed libration amplitudes in the 5:2 is likely inconsistent with transient sticking.

In the future, our results can be used to compare with well-characterized surveys such as OSSOS to constrain the transient-sticking component of TNO populations in mean motion resonance with Neptune.  Libration amplitudes provide a useful dynamical tracer of this population, which in conjunction with physical properties consistent with those of the scattering population, may be used to identify transient interlopers even in resonances dominated by other emplacement mechanisms.  Our initial comparison is already quite interesting:  the distant 5:2 population has a large population that appears inconsistent---both in number and in libration amplitude distribution---with sole emplacement by transient sticking from the \textit{current} scattered disk.  Models of dynamical sculpting of the outer solar system will need to account for this 5:2 population in another way. In other words, we find that a large fraction of the 5:2 resonant population must have been emplaced early in the life of the solar system.

\acknowledgements
KV and RMC acknowledge support from NASA Solar System Workings grant NNX15AH59G. KV acknowledges additional support from NASA grant NNX14AG93G.
We thank Sarah Greenstreet for providing the scattered disk initial conditions used in this work and Nate Kaib for the simulation data from which those initial conditions were derived.  We thank Patryk Lykawka for helpful discussions during the course of this work.

\clearpage


\begin{thebibliography}{38}
\expandafter\ifx\csname natexlab\endcsname\relax\def\natexlab#1{#1}\fi

\bibitem[{{Adams} {et~al.}(2014){Adams}, {Gulbis}, {Elliot}, {Benecchi},
  {Buie}, {Trilling}, \& {Wasserman}}]{Adams2014}
{Adams}, E.~R., {Gulbis}, A.~A.~S., {Elliot}, J.~L., {Benecchi}, S.~D., {Buie},
  M.~W., {Trilling}, D.~E., \& {Wasserman}, L.~H. 2014, \aj, 148, 55

\bibitem[{{Alexandersen} {et~al.}(2013){Alexandersen}, {Gladman},
  {Greenstreet}, {Kavelaars}, {Petit}, \& {Gwyn}}]{Alexandersen2013}
{Alexandersen}, M., {Gladman}, B., {Greenstreet}, S., {Kavelaars}, J.~J.,
  {Petit}, J.-M., \& {Gwyn}, S. 2013, Science, 341, 994

\bibitem[{Alexandersen {et~al.}(2016)Alexandersen, Gladman, Kavelaars, Petit,
  Gwyn, Shankman, \& Pike}]{Alexandersen2016}
Alexandersen, M., Gladman, B., Kavelaars, J.~J., Petit, J.-M., Gwyn, S. D.~J.,
  Shankman, C.~J., \& Pike, R.~E. 2016, \aj, 152, 1

\bibitem[{{Bannister} {et~al.}(2016{\natexlab{a}}){Bannister}, {Alexandersen},
  {Benecchi}, {Chen}, {Delsanti}, {Fraser}, {Gladman}, {Granvik}, {Grundy},
  {Guilbert-Lepoutre}, {Gwyn}, {Ip}, {Jakubik}, {Jones}, {Kaib}, {Kavelaars},
  {Lacerda}, {Lawler}, {Lehner}, {Lin}, {Lykawka}, {Marsset}, {Murray-Clay},
  {Noll}, {Parker}, {Petit}, {Pike}, {Rousselot}, {Schwamb}, {Shankman},
  {Veres}, {Vernazza}, {Volk}, {Wang}, \& {Weryk}}]{Bannister201692}
{Bannister}, M.~T., {et~al.} 2016{\natexlab{a}}, \aj, 152, 212

\bibitem[{{Bannister} {et~al.}(2016{\natexlab{b}}){Bannister}, {Kavelaars},
  {Petit}, {Gladman}, {Gwyn}, {Chen}, {Volk}, {Alexandersen}, {Benecchi},
  {Delsanti}, {Fraser}, {Granvik}, {Grundy}, {Guilbert-Lepoutre}, {Hestroffer},
  {Ip}, {Jakubik}, {Jones}, {Kaib}, {Lacerda}, {Lawler}, {Lehner}, {Lin},
  {Lister}, {Lykawka}, {Monty}, {Marsset}, {Murray-Clay}, {Noll}, {Parker},
  {Pike}, {Rousselot}, {Rusk}, {Schwamb}, {Shankman}, {Sicardy}, {Vernazza}, \&
  {Wang}}]{Bannister2016}
---. 2016{\natexlab{b}}, \aj, 152, 70

\bibitem[{{Bannister} {et~al.}(2018){Bannister}, {Kavelaars}, {Petit},
  {Gladman}, {Gwyn}, {Chen}, {Volk}, {Alexandersen}, {Benecchi}, {Delsanti},
  {Fraser}, {Granvik}, {Grundy}, {Guilbert-Lepoutre}, {Hestroffer}, {Ip},
  {Jakubik}, {Jones}, {Kaib}, {Lacerda}, {Lawler}, {Lehner}, {Lin}, {Lister},
  {Lykawka}, {Monty}, {Marsset}, {Murray-Clay}, {Noll}, {Parker}, {Pike},
  {Rousselot}, {Rusk}, {Schwamb}, {Shankman}, {Sicardy}, {Vernazza}, \&
  {Wang}}]{Bannister2018}
---. 2018, ApJS in press

\bibitem[{{Beauge}(1994)}]{Beauge1994}
{Beauge}, C. 1994, Celestial Mechanics and Dynamical Astronomy, 60, 225

\bibitem[{{Chiang} {et~al.}(2003){Chiang}, {Jordan}, {Millis}, {Buie},
  {Wasserman}, {Elliot}, {Kern}, {Trilling}, {Meech}, \& {Wagner}}]{Chiang2003}
{Chiang}, E.~I., {et~al.} 2003, \aj, 126, 430

\bibitem[{{Dawson} \& {Murray-Clay}(2012)}]{Dawson2012}
{Dawson}, R.~I., \& {Murray-Clay}, R. 2012, \apj, 750, 43

\bibitem[{{Dones} {et~al.}(2004){Dones}, {Weissman}, {Levison}, \&
  {Duncan}}]{Dones2004}
{Dones}, L., {Weissman}, P.~R., {Levison}, H.~F., \& {Duncan}, M.~J. 2004,
  {Oort cloud formation and dynamics}, ed. G.~W. {Kronk}, 153--174

\bibitem[{{Duncan} \& {Levison}(1997)}]{Levison1997}
{Duncan}, M.~J., \& {Levison}, H.~F. 1997, Science, 276, 1670

\bibitem[{{Gladman} {et~al.}(2008){Gladman}, {Marsden}, \&
  {Vanlaerhoven}}]{Gladman2008}
{Gladman}, B., {Marsden}, B.~G., \& {Vanlaerhoven}, C. 2008, {Nomenclature in
  the Outer Solar System}, ed. M.~A. {Barucci}, H.~{Boehnhardt}, D.~P.
  {Cruikshank}, A.~{Morbidelli}, \& R.~{Dotson}, 43--57

\bibitem[{{Gladman} {et~al.}(2012){Gladman}, {Lawler}, {Petit}, {Kavelaars},
  {Jones}, {Parker}, {Van Laerhoven}, {Nicholson}, {Rousselot}, {Bieryla}, \&
  {Ashby}}]{Gladman2012}
{Gladman}, B., {et~al.} 2012, \aj, 144, 23

\bibitem[{{Gomes} {et~al.}(2008){Gomes}, {Fern Ndez}, {Gallardo}, \&
  {Brunini}}]{Gomes2008}
{Gomes}, R.~S., {Fern Ndez}, J.~A., {Gallardo}, T., \& {Brunini}, A. 2008, 259

\bibitem[{{Gulbis} {et~al.}(2010){Gulbis}, {Elliot}, {Adams}, {Benecchi},
  {Buie}, {Trilling}, \& {Wasserman}}]{Gulbis2010}
{Gulbis}, A.~A.~S., {Elliot}, J.~L., {Adams}, E.~R., {Benecchi}, S.~D., {Buie},
  M.~W., {Trilling}, D.~E., \& {Wasserman}, L.~H. 2010, \aj, 140, 350

\bibitem[{{Hahn} \& {Malhotra}(2005)}]{Hahn2005}
{Hahn}, J.~M., \& {Malhotra}, R. 2005, \aj, 130, 2392

\bibitem[{{Holman} {et~al.}(2018){Holman}, {Payne}, {Fraser}, {Lacerda},
  {Bannister}, {Lackner}, {Chen}, {Lin}, {Smith}, {Kokotanekova}, {Young},
  {Chambers}, {Chastel}, {Denneau}, {Fitzsimmons}, {Flewelling}, {Grav},
  {Huber}, {Induni}, {Kudritzki}, {Krolewski}, {Jedicke}, {Kaiser}, {Lilly},
  {Magnier}, {Mark}, {Meech}, {Micheli}, {Murray}, {Parker}, {Protopapas},
  {Ragozzine}, {Veres}, {Wainscoat}, {Waters}, \& {Weryk}}]{Holman2018}
{Holman}, M.~J., {et~al.} 2018, \apjl, 855, L6

\bibitem[{{Kaib} {et~al.}(2011){Kaib}, {Ro{\v s}kar}, \& {Quinn}}]{Kaib2011}
{Kaib}, N.~A., {Ro{\v s}kar}, R., \& {Quinn}, T. 2011, \icarus, 215, 491

\bibitem[{{Kaib} \& {Sheppard}(2016)}]{Kaib2016}
{Kaib}, N.~A., \& {Sheppard}, S.~S. 2016, \aj, 152, 133

\bibitem[{{Lawler} {et~al.}(2018){Lawler}, {Shankman}, {Kavelaars},
  {Alexandersen}, {Bannister}, {Chen}, {Gladman}, {Fraser}, {Gwyn}, {Kaib},
  {Petit}, \& {Volk}}]{Lawler2018}
{Lawler}, S.~M., {et~al.} 2018, ArXiv e-prints

\bibitem[{{Levison} \& {Duncan}(1994)}]{Levison1994}
{Levison}, H.~F., \& {Duncan}, M.~J. 1994, \icarus, 108, 18

\bibitem[{{Levison} {et~al.}(2008){Levison}, {Morbidelli}, {Vanlaerhoven},
  {Gomes}, \& {Tsiganis}}]{Levison2008}
{Levison}, H.~F., {Morbidelli}, A., {Vanlaerhoven}, C., {Gomes}, R., \&
  {Tsiganis}, K. 2008, \icarus, 196, 258

\bibitem[{{Lykawka} \& {Mukai}(2007)}]{Lykawka2007}
{Lykawka}, P.~S., \& {Mukai}, T. 2007, \icarus, 192, 238

\bibitem[{{Malhotra}(1993)}]{Malhotra1993}
{Malhotra}, R. 1993, \nat, 365, 819

\bibitem[{{Malhotra}(1995)}]{Malhotra1995}
---. 1995, \aj, 110, 420

\bibitem[{{Malhotra} {et~al.}(2018){Malhotra}, {Lan}, {Volk}, \&
  {Wang}}]{Malhotra:2018}
{Malhotra}, R., {Lan}, L., {Volk}, K., \& {Wang}, X. 2018, ArXiv e-prints

\bibitem[{{Morbidelli} {et~al.}(2008){Morbidelli}, {Levison}, \&
  {Gomes}}]{Morbidelli2008}
{Morbidelli}, A., {Levison}, H.~F., \& {Gomes}, R. 2008, {The Dynamical
  Structure of the Kuiper Belt and Its Primordial Origin}, ed. {Barucci, M.~A.,
  Boehnhardt, H., Cruikshank, D.~P., Morbidelli, A., \& Dotson, R.}, 275--292

\bibitem[{{Murray} \& {Dermott}(1999)}]{MurrayDermott1999}
{Murray}, C.~D., \& {Dermott}, S.~F. 1999, {Solar system dynamics} (Cambridge:
  University Press)

\bibitem[{{Nesvorn{\'y}} \& {Roig}(2001)}]{Nesvorny2001}
{Nesvorn{\'y}}, D., \& {Roig}, F. 2001, \icarus, 150, 104

\bibitem[{{Nesvorn{\'y}} {et~al.}(2016){Nesvorn{\'y}}, {Vokrouhlick{\'y}}, \&
  {Roig}}]{Nesvorny2016}
{Nesvorn{\'y}}, D., {Vokrouhlick{\'y}}, D., \& {Roig}, F. 2016, \apjl, 827, L35

\bibitem[{{Petit} {et~al.}(2011){Petit}, {Kavelaars}, {Gladman}, {Jones},
  {Parker}, {Van Laerhoven}, {Nicholson}, {Mars}, {Rousselot}, {Mousis},
  {Marsden}, {Bieryla}, {Taylor}, {Ashby}, {Benavidez}, {Campo Bagatin}, \&
  {Bernabeu}}]{Petit2011}
{Petit}, J.-M., {et~al.} 2011, \aj, 142, 131

\bibitem[{{Pike} {et~al.}(2015){Pike}, {Kavelaars}, {Petit}, {Gladman},
  {Alexandersen}, {Volk}, \& {Shankman}}]{Pike2015}
{Pike}, R.~E., {Kavelaars}, J.~J., {Petit}, J.~M., {Gladman}, B.~J.,
  {Alexandersen}, M., {Volk}, K., \& {Shankman}, C.~J. 2015, \aj, 149, 202

\bibitem[{{Pike} {et~al.}(2017){Pike}, {Lawler}, {Brasser}, {Shankman},
  {Alexandersen}, \& {Kavelaars}}]{Pike2017}
{Pike}, R.~E., {Lawler}, S., {Brasser}, R., {Shankman}, C.~J., {Alexandersen},
  M., \& {Kavelaars}, J.~J. 2017, \aj, 153, 127

\bibitem[{{Shankman} {et~al.}(2013){Shankman}, {Gladman}, {Kaib}, {Kavelaars},
  \& {Petit}}]{Shankman2013}
{Shankman}, C., {Gladman}, B.~J., {Kaib}, N., {Kavelaars}, J.~J., \& {Petit},
  J.~M. 2013, \apjl, 764, L2

\bibitem[{Shankman {et~al.}(2016)Shankman, Kavelaars, Gladman, Alexandersen,
  Kaib, Petit, Bannister, Chen, Gwyn, Jakub{\'\i}k, \& Volk}]{Shankman2016}
Shankman, C., {et~al.} 2016, The Astronomical Journal, 151, 1

\bibitem[{{Volk} {et~al.}(2016){Volk}, {Murray-Clay}, {Gladman}, {Samantha
  Lawler}, {Michele T. Bannister}, {J. J. Kavelaars}, {Jean-Marc Petit},
  {Stephen Gwyn}, {Mike Alexandersen}, {Ying-Tung Chen}, {Patryk Sofia
  Lykawka}, {Wing Ip}, \& {Hsing Wen Lin}}]{Volk2016}
{Volk}, K., {et~al.} 2016, \aj, 152, 23

\bibitem[{{Volk} {et~al.}(2018){Volk}, {Murray-Clay}, {Gladman}, {Lawler},
  {Yu}, {Alexandersen}, {Bannister}, {Chen}, {Dawson}, {Greenstreet}, {Gwyn},
  {Kavelaars}, {Lin}, {Lykawka}, \& {Petit}}]{Volk:2018}
---. 2018, ArXiv e-prints

\bibitem[{{Wang} \& {Malhotra}(2017)}]{Wang:2017}
{Wang}, X., \& {Malhotra}, R. 2017, \aj, 154, 20

\end{thebibliography}
\end{document}